\documentclass[aps,pra,twocolumn,groupedaddress,nofootinbib,notitlepage,showpacs,floatfix]{revtex4}

\usepackage{graphicx,graphics,epsfig,times,bm,,bbm,amssymb,amsmath,amsfonts,mathrsfs,color}
\usepackage{hypernat}
\usepackage{hyperref}
\usepackage{amsthm,mathtools,natbib,multirow,epsf,latexsym,setspace,array, xcolor}
\usepackage[mmddyy]{datetime}

\newtheorem{mylemma}{Lemma}

\newcommand{\ignore}[1]{}

\newtheorem{QDD}{QDD Theorem}
\newtheorem{mydef}{Definition}

\def\s0{\openone}
\def\sx{\sigma_X}
\def\sy{\sigma_Y}
\def\sz{\sigma_Z}
\newcommand{\red}[1]{\textcolor{black}{#1}}
\newcommand{\blue}[1]{\textcolor{black}{#1}}

\newcommand{\beq} {\begin{equation}}
\newcommand{\eeq} {\end{equation}}
\newcommand{\bea} {\begin{eqnarray}}
\newcommand{\eea} {\end{eqnarray}}
\newcommand{\bes} {\begin{subequations}}
\newcommand{\ees} {\end{subequations}}

\begin{document}

\title{Quadratic Dynamical Decoupling: Universality Proof and Error Analysis}
\author{Wan-Jung Kuo}
\email{wanjungk@usc.edu}
\affiliation{Department of Physics and Center for Quantum Information Science \&
Technology, University of Southern California, Los Angeles, California
90089, USA}
\author{Daniel A. Lidar}
\email{lidar@usc.edu}
\affiliation{Departments of Electrical Engineering, Chemistry, and Physics, and Center
for Quantum Information Science \& Technology, University of Southern
California, Los Angeles, California 90089, USA}

\begin{abstract}
We prove the universality of the generalized QDD$_{N_1 N_2}$ (quadratic dynamical decoupling) pulse sequence for near-optimal suppression of general single-qubit decoherence. Earlier work showed numerically that this dynamical decoupling sequence, which consists of an inner Uhrig DD (UDD) and outer UDD sequence using $N_1$ and $N_2$ pulses respectively, can eliminate decoherence to $\mathcal{O}(T^N)$ using $\mathcal{O}(N^2)$ unequally spaced ``ideal" (zero-width) pulses, where $T$ is the total evolution time and $N=N_1=N_2$. A proof of the universality of QDD has been given for even $N_1$. Here we give a general universality proof of QDD for arbitrary $N_1$ and $N_2$. As in earlier proofs, our result holds for arbitrary bounded environments. Furthermore, we explore the single-axis (polarization) error suppression abilities of the inner and outer UDD sequences. We analyze both the single-axis QDD performance and how the overall performance of QDD depends on the single-axis errors.  We identify various performance effects related to the parities and relative magnitudes of $N_1$ and $N_2$. We prove that using QDD$_{N_1 N_2}$ decoherence can always be eliminated to $\mathcal{O}(T^{\min\{N_1,N_2\}})$.
\end{abstract}
\ignore{
We prove the universality of the generalized QDD_{N1,N2} (quadratic dynamical decoupling) pulse sequence for near-optimal suppression of general single-qubit decoherence. Earlier work showed numerically that this dynamical decoupling sequence, which consists of an inner Uhrig DD (UDD) and outer UDD sequence using N1 and N2 pulses respectively, can eliminate decoherence to O(T^N) using O(N^2) unequally spaced "ideal" (zero-width) pulses, where T is the total evolution time and N=N1=N2. A proof of the universality of QDD has been given for even N1. Here we give a general universality proof of QDD for arbitrary N1 and N2. As in earlier proofs, our result holds for arbitrary bounded environments. Furthermore, we explore the single-axis (polarization) error suppression abilities of the inner and outer UDD sequences. We analyze both the single-axis QDD performance and how the overall performance of QDD depends on the single-axis errors.  We identify various performance effects related to the parities and relative magnitudes of N1 and N2. We prove that using QDD_{N1,N2} decoherence can always be eliminated to O(T^min[N1,N2]).
}

\maketitle

\section{Introduction}
\label{intro}

The inevitable coupling between a quantum system and its environment, or bath, typically results in decoherence \cite{Breuer:book}. It is essential in quantum information processing (QIP) to find protection against decoherence, as it leads to computational errors which can quickly eliminate any quantum advantage \cite{NielsenChuang:book,Ladd:10}.  A powerful technique that can be used to this end, adapted from nuclear magnetic resonance (NMR) refocusing techniques developed since the discovery of the spin echo effect \cite{Hahn:50,Haeberlen:book}, is dynamical decoupling (DD) \cite{Yang-DD-review}.
It mitigates the unwanted system-bath interactions through the application of a sequence of short and strong pulses, acting purely on the system.  DD is an open-loop technique which works when the bath is non-Markovian \cite{Breuer:book}, and bypasses the need for measurement or feedback, in contrast to closed-loop quantum error correction (QEC) \cite{Gottesman:10}. However, while QEC can be made fault tolerant \cite{TerhalBurkard:05,Aliferis:05}, it is unlikely that this holds for DD as a stand-alone method, or that it holds for any other purely open-loop method, for that matter \cite{PhysRevA.83.020305}. This notwithstanding, DD can significantly improve the performance of fault tolerant QEC when the two methods are combined \cite{NLP:09}.

DD was first introduced into QIP in order to preserve single-qubit coherence within the spin-boson model \cite{ViolaLloyd:98,Ban:98,Duan:98e}. It was soon generalized via a dynamical group symmetrization framework to preserving the states of open quantum systems interacting with arbitrary (but bounded) environments \cite{Zanardi:99,ViolaKnillLloyd:99}. These early DD schemes work to a given low order in time-dependent perturbation theory (e.g., the Magnus or Dyson expansions \cite{Blanes:08}). Namely, the effective system-bath interaction following a DD pulse sequence lasting for a  total time $T$ only contains terms of order $T^{N+1}$ and higher, where typically $N$ was $1$ for the early DD schemes. For general $N$ this is called $N^{\rm th}$-order decoupling. Concatenated DD (CDD) \cite{KhodjastehLidar:05}, where a given pulse sequence is recursively embedded into itself, was the first explicit scheme capable of achieving arbitrary order decoupling, i.e., CDD allows $N$ to be tuned at will \cite{KhodjastehLidar:07}. CDD has been amply tested in recent experimental studies \cite{Peng:11,Alvarez:10,2010arXiv1011.1903T,2010arXiv1011.6417W,2010arXiv1007.4255B}, and demonstrated to be fairly robust against pulse imperfections. However, the number of pulses CDD requires grows exponentially with $N$. In order to implement scalable QIP
it is desirable to design efficient DD schemes which have as few pulses as possible. 

For the one qubit pure dephasing spin-boson model, Uhrig discovered a DD sequence (UDD) which is optimal in the sense that it achieves $N$th order decoupling with the smallest possible number of pulses, $N$ or $N+1$, depending on whether $N$ is even or odd \cite{Uhrig:07}. The key difference compared to other DD schemes is that in UDD the pulses are applied at non-uniform intervals. This optimal pulse sequence had also been noticed in \cite{DharGroverRoy:06} for $N \leq 5$. A scheme to protect a known two-qubit entangled state using UDD was given in Ref.~\cite{Mukhtar:10}. UDD was conjectured to be model-independent (``universal") with an analytical verification up to $N=9$ \cite{LeeWitzelDasSarma:08} and $N=14$ \cite{Uhrig:08}. A general proof of universality of the UDD sequence was first given in \cite{YangLiu:08} (see also Ref.~\cite{UL:10} for an alternative proof). The performance of the UDD sequence was the subject of a wide range of recent experimental studies \cite{Biercuk:09,Biercuk2:09,Du:09,Alvarez:10,Elizabeth:09,Barthel:10}. An interesting application was to the enhancement of magnetic resonance imaging of structured materials such as tissue \cite{Jenista2009Optimized}. However, one conclusion from some of these studies is that the superior convergence of UDD compared to CDD comes at the expense of lack of robustness to pulse imperfections. It is possible that recent theoretical pulse shaping developments \cite{PasiniFischerKarbachUhrig:08,UhrigPasini:10}, designed to replace ideal, instantaneous pulses with realistic pulses of finite duration and amplitude while maintaining the suppression properties of UDD, will lead to improved experimental robustness.

The UDD sequence is effective not only against pure dephasing but also against longitudinal relaxation of a qubit coupled to an arbitrary bounded environment \cite{YangLiu:08}. That is, UDD efficiently suppresses pairs of single-axis errors. However, it cannot overcome general, three-axis qubit decoherence. The reason is that UDD uses a single pulse type (e.g., pulses along the $x$-axis of the qubit Bloch sphere), and system-bath interactions which commute with this pulse type are unaffected by the sequence.

Combining orthogonal single-axis CDD and UDD sequences (CUDD) reduces the number of control pulses required for the suppression of general single-qubit decoherence compared to two-axis CDD \cite{Uhrig:09}. However, CUDD still requires an exponential number of pulses. This scaling problem was overcome with the introduction of the quadratic DD (QDD) sequence by West {\it et al.}, which nests two UDD sequences with different pulse types and different numbers of pulses $N_1$ and $N_2$ \cite{WestFongLidar:10}. We denote this sequence by QDD$_{N_1 N_2}$, where $N_1$ and $N_2$ are the numbers of pulses of the inner and outer UDD sequences, respectively. QDD$_{N,N}$ (where the inner and outer UDD sequences have the same decoupling order) was conjectured to  suppress arbitrary qubit-bath coupling to order $N$ by using $\mathcal{O}(N^{2})$ pulses, an exponential improvement over all previously known DD schemes for general qubit decoherence \cite{WestFongLidar:10}. This conjecture was based on numerical studies for $N\leq 6$ \cite{WestFongLidar:10}, and these were recently extended to $N\leq 24$ \cite{QuirozLidar:11}, in support of the conjecture. An early argument for the universality and performance of QDD (which below we refer to as ``validity of QDD"), based on an extension of UDD to analytically time-dependent Hamiltonians \cite{PasiniUhrig:10}, fell short of a proof since the effective Hamiltonian resulting from the inner UDD sequences in QDD is not analytic.

The problem of finding a proof of the validity of QDD was first successfully addressed by Wang \& Liu \cite{WangLiu:11}, though not in complete generality, as we explain below. In fact Ref.~\cite{WangLiu:11} considered the more general problem of protecting a set of qubits or multilevel systems against arbitrary system-bath interactions, using a nested UDD (NUDD) scheme, a generalization of QDD to multiple nested UDD sequences. This problem was also studied, for two qubits, by Mukhtar {\it et al}., \red{whose numerical results} showed, for their specific choice of pulse operators, that the ordering of the nested UDD sequences impacts performance \cite{Mukhtar2:10}.  Wang \& Liu's proof is based on the idea of using mutually
orthogonal operation (MOOS) sets---mutually commuting or anti-commuting unitary Hermitian system operators---as control pulses \cite{WangLiu:11} (the ordering effect observed in Ref.~\cite{Mukhtar2:10} disappears when using MOOS sets). As in QDD, the decoupling orders of the nested UDD sequences in NUDD can be different, so that different error types can be removed to different orders. Wang \& Liu proved the validity of the general QDD/NUDD scheme when the order of all inner UDD sequences is {\em even} (the order of the outermost sequence can be even or odd) \cite{WangLiu:11}. Their proof is based on MOOS set preservation, and does not apply to \red{QDD, or more generally NUDD, when the order of at least one of the inner UDD sequences is odd}. In addition, Wang \& Liu pointed out that there are QDD$_{N_1 N_2}$ examples showing that the outer level UDD sequence does not work ``as expected" (i.e., does not suppress errors to its order) if $N_1$ is odd and $N_1 < N_2$. Thus their proof left the actual suppression order of QDD/NUDD with odd order UDD at the inner levels as an open question.

This problem was addressed numerically in Ref.~\cite{QuirozLidar:11}, which studied the performance of QDD$_{N_1 N_2}$ for all three single-axis errors. The numerical results show that the suppression ability of the outer UDD sequence is indeed hindered by the inner UDD sequence if $N_1$ is odd and, surprisingly, smaller than \emph{half} of the order of the outer level UDD sequence. Moreover, Ref.~\cite{QuirozLidar:11} reported that the suppression order of the system-bath interaction which anti-commutes with the pulses of both the inner and outer sequences depends on the parities of both $N_1$ and $N_2$. 

In this work we provide a complete proof of the validity of QDD$_{N_1 N_2}$. In particular, we also prove the case of odd $N_1$ left open in Ref.~\cite{WangLiu:11}. Moreover, we analyze the single-axis error suppression abilities of both the inner and outer UDD sequences, and thus provide analytical bounds in support of the numerical results of Ref.~\cite{QuirozLidar:11}.

We show that the single-axis error which anti-commutes with the pulses of the inner sequence but commutes with those of the outer sequence is always suppressed to the expected order ($N_1$). The suppression of the two other single-axis errors (the one which commutes with the inner sequence pulses but anti-commutes with the outer sequence pulses, and the one which anti-commutes with both), is more subtle, and depends on the relative size and parity of $N_1$ and $N_2$.

Specifically, we show that when $N_1$ is even, QDD$_{N_1 N_2}$ always achieves at least the expected decoupling order, irrespective of the relative size of $N_1$ and $N_2$. However, when $N_1$ is odd and $N_1 < N_2-1$, we show that  the decoupling order of the error which commutes with the inner sequence pulses but anti-commutes with those of the outer sequence, is at least $N_1+1$, smaller than the expected suppression order ($N_2$). 
\ignore{Thus, \red{an odd-order  inner sequence can prevent the outer sequence from reaching its expected suppression order.}\blue{I feel that using "can" is too strong. Our method shows the lower bound of decoupling order of error Z is $N_1+1$ but doesn't show the decoupling order is exactly and not higher than $N_1+1$.  So I feel that it's better to say that  it's likely that inner sequence can hinder the outer sequence.}
}
Nevertheless, for odd $N_1$ and $N_1 \geq N_2-1$, the outer UDD sequence always suppresses the error which commutes with the inner sequence pulses but anti-commutes with those of the outer sequence to the expected order ($N_2$).
\ignore{we show that the inner  sequence prevents the outer sequence from reaching its expected suppression order ($N_2$). Consequently the decoupling order of the error which commutes with the inner sequence pulses but anti-commutes with those of the outer sequence, is only $N_1+1$.  
This can be interpreted as being the outcome of having an inner sequence which leaves residual errors which cannot be corrected by the outer sequence, since the latter's pulses commute with these errors. As a result the bottleneck is the inner sequence.}

One might expect that the error which anti-commutes with the pulses of both the inner and outer sequences can be suppressed by both sequences. In other words, one might expect this error to be removed at least up to order $\max[N_1,N_2]$. However, we show that this expectation is fulfilled only when $N_1$ is even.  When $N_1$ is odd, it determines the suppression order. However, interestingly, the parity of $N_{2}$ also plays a role, namely, when it is odd the suppression order is one order higher than when $N_2$ is even.

Despite this complicated interplay between the orders of the inner and outer UDD sequences, resulting in the outer sequence not always achieving its expected decoupling order when $N_1$ is odd, we show that, overall, QDD$_{N_1 N_2}$ always suppresses all single-qubit errors at least to order $\min[N_1,N_2]$.

A complete summary of our results for the different single-axis suppression orders under QDD$_{N_1 N_2}$ is given in Table \ref{table:qdd}. Our analytical results are in complete agreement with the numerical findings of Ref.~\cite{QuirozLidar:11}, but our proof method underestimates the suppression of of the error which commutes with the inner sequence pulses but anti-commutes with those of the outer sequence: for odd $N_1$ we find a decoupling order of $\min[N_1+1,N_2]$, while the numerical result is $\min[2N_1+1,N_2]$ for $N_1,N_2 \leq 24$. Explaining this discrepancy is thus still an open problem.

The structure of the paper is as follows. The model of general decoherence of one qubit in the presence of instantaneous QDD pulses is defined in Sec.~\ref{sec:model}. The QDD theorem is stated there as well. We prove the QDD theorem in Sec.~\ref{sec: ax and ay} and Sec.~\ref{sec:outer}. A comparison between the numerical results of Ref.~\cite{QuirozLidar:11} and our theoretical bounds is presented in Sec.~\ref{sec: numerical and theory}. We conclude in Sec.~\ref{sec:conclusion}. The appendix provide additional technical details.


\section{\protect System-bath model and the QDD sequence}
\label{sec:model}

\subsection{General QDD$_{N_{1} N_{2}}$ scheme}

We model general decoherence on a single qubit via the following Hamiltonian:
\begin{equation}
H=J_{0}I\otimes B_{0}+J_{X}\sigma _{X}\otimes B_{X}+J_{Y}\sigma _{Y}\otimes
B_{Y}+J_{Z}\sigma _{Z}\otimes B_{Z},
\label{H}
\end{equation}%
where $B_{\lambda}$, $\lambda \in \{0,X,Y,Z\}$, are arbitrary bath-operators with $\| B_{\lambda}\|=1$ (the norm is the largest singular value), the Pauli matrices, $\sigma _{\lambda}$, $\lambda \in \{X,Y,Z\}$, are the unwanted errors acting on the system qubit, and $J_{\lambda}$, $\lambda \in \{0,X,Y,Z\}$, are bounded coupling coefficients between the qubit and the bath. 

The QDD$_{N_{1} N_{2}}$ pulse sequence is constructed by nesting a $Z$-type UDD$_{N_{1}}$ sequence, designed to eliminate the longitudinal relaxation errors $\sigma _{X}\otimes B_{X}$ and $\sigma _{Y}\otimes B_{Y}$ up to order $T^{N_{1}+1}$ by using $N_{1}$ or $N_{1}+1$ pulses, with an $X$-type UDD$_{N_{2}}$ sequence, designed to eliminate the pure dephasing error $\sigma _{Z}\otimes B_{Z}$ up to order $T^{N_{2}+1}$ by using $N_{2}$ or $N_{2}+1$ pulses. The nesting order does not matter for our analysis, so without loss of generality we choose $Z$-type UDD$_{N_{1}}$ to be the inner sequence and $X$-type UDD$_{N_{2}}$ to be the outer sequence.  We use the notation $X$ and $Z$ to denote control pulses, to distinguish the same operators from the unwanted errors denoted by the Pauli matrices. We also sometimes use the notation $\sigma _{0}$ for the $2\times 2$ identity matrix $I$.

The $X$-type UDD$_{N_{2}}$ pulses comprising the outer layer of the QDD$_{N_{1} N_{2}}$ sequence are applied at the original UDD$_{N_{2}}$ timing with total evolution time $T$, $t_{j}=T\eta_{j}$ where $\eta_{j}$ is the normalized UDD timing (or the normalized QDD$_{N_{1} N_{2}}$ outer sequence timing),
\beq
\eta_{j}=\sin ^{2}\frac{j\pi }{2(N_{2}+1)}
\label{udd time}
\eeq
with $j=1,2,\dots , \overline{N}_{2}$ where $\overline{N}_{2}=N_{2}$ if $N_{2}$ even and $\overline{N}_{2}=N_{2}+1$ if $N_{2}$ odd. The additional pulse applied at the end of the sequence when $N_{2}$ is odd, is required in order to make the total number of $X$ pulses-type even, so that the overall effect of the $X$-type pulses at the final time $T$ will be to leave the qubit state unchanged. Note that it is the relative size of the pulse intervals that matters for error cancellations in UDD, not the precise pulse application times. Hence, the most relevant quantities for the outer level  UDD$_{N_{2}}$ are  the $N_{2}+1$ normalized UDD$_{N_{2}}$ pulse intervals (or the normalized QDD$_{N_{1} N_{2}}$ outer pulse intervals),  
\bes
\bea
s_{j}&\equiv& \frac{\tau _{j}}{T}=\eta_j-\eta_{j-1} \label{sj} \\
&=& \sin \frac{\pi }{2(N_{2}+1)}\sin \frac{(2j-1)\pi }{2(N_{2}+1)} 
\eea
\ees
where $\tau _{j}\equiv t_{j}-t_{j-1} $ is the actual pulse interval.

\begin{table}
\begin{tabular}{|l|c|c|c|c|}
\hline
 & \multicolumn{2}{c|}{inner order $N_1$} & \multicolumn{2}{c|}{outer order $N_2$}  \\ 
 & even & odd & even & odd     \\ \hline
no. of pulses  & $N_1$ & $N_1+1$ & $N_2$ & $N_1+1$     \\ \hline
no. of intervals & $N_1+1$ & $N_1+1$ & $N_2+1$ & $N_2+1$   \\ \hline
\end{tabular}
\caption{\red{Inner and outer sequence orders $N_1$ and $N_2$ {\it vs} the number of pulses and pulse intervals in the inner and outer sequences.}}
\label{tab:pulses}\centering
\end{table}

The $Z$-type pulses of the inner level UDD$_{N_{1}}$, applied from $t_{j-1}$ to $t_{j}$, are executed at times 
\beq
t_{j,k}=t_{j-1}+\tau _{j}\sin ^{2}\frac{k\pi }{2(N_{1}+1)}
\eeq
with $N_{1}+1$ pulse intervals 
\beq
\tau _{j,k} \equiv t_{j,k}-t_{j,k-1}.
\eeq
Even though adding an additional $Z$-type pulse to the end of each inner sequence with odd $N_{1}$ is not required (since instead one can add just one additional $Z$-type pulse at the end of the QDD$_{N_{1} N_{2}}$ sequence to ensure that the total number of $Z$ pulses at the final time $T$ is even), 
for simplicity of our later analysis, we let $k=1,2,\dots \overline{N}_{1}$ where $\overline{N}_{1}=N_{1}$ if $N_{1}$ even and $\overline{N}_{1}=N_{1}+1$ if $N_{1}$ odd. The corresponding normalized QDD$_{N_{1} N_{2}}$ inner pulse timings are
\begin{equation}
\eta_{j,k} \equiv \frac{t_{j,k}}{T}=\eta_{j-1}+s _{j}\sin ^{2}\frac{k\pi }{2(N_{1}+1)} 
\label{eta-jk}
\end{equation}%
with the normalized QDD$_{N_{1} N_{2}}$ inner pulse interval $\frac{\tau_{j,k}}{T} =s_{j}\tilde{s}_{k}$,  where $\tilde{s}_{k}$ is the normalized UDD$_{N_{1}}$ pulse interval and is the same function as $s_{j}$ but with different decoupling order $N_{1}$. The first subindex stands for the outer interval while the second subindex stands for the inner interval.  Moreover, by definition, we have 
\beq
\eta_{j}=\eta_{j,N_{1}+1}=\eta_{j+1,0}.
\eeq

To summarize, the evolution operator at the final time $T$, at the completion of the QDD$_{N_{1} N_{2}}$ sequence, is
\begin{equation}
U(T)=X^{N_{2}}U_{Z}(\tau _{N_{2}+1})X\cdots XU_{Z}(\tau _{2})XU_{Z}(\tau _{1}),
\label{qddsequence}
\end{equation}%
with 
\begin{equation}
U_{Z}(\tau _{j})=Z^{N_{1}}U_{f}(\tau _{j,N_{1}+1})Z\cdots ZU_{f}(\tau _{j,2})ZU_{f}(\tau _{j,1})
\end{equation}%
being the inner UDD$_{N_{1}}$ sequence evolution, and with $U_{f}$ being the pulse-free evolution generated by $H$ [Eq.~\eqref{H}].  Table~\ref{tab:pulses} summarizes \red{how the number of pulses and pulse intervals in the inner and outer sequences depend on the inner and outer sequence orders $N_1$ and $N_2$.}


\subsection{Toggling frame}

Our QDD proof will be done in the toggling frame. Since our analysis is based on an expansion of powers of the total time $T$, most quantities we will deal with are functions of the normalized total time $1$.  

The normalized control Hamiltonian with $\eta\equiv\frac{t}{T}$ is given by, 
\begin{equation}
H_{c}(\eta)=\frac{\pi }{2}[X\sum_{j=1}^{\overline{N}_{2}}\delta(\eta-\eta_{j})+Z\sum_{j=1}^{N_{2}+1}\sum_{k=1}^{\overline{N}_{1}}\delta(\eta-\eta_{j,k})].
\end{equation}

The normalized control evolution operator, 
\begin{eqnarray}
U_{c}(\eta)=\widehat{T}\exp [-i\int_{0}^{\eta}H_{c}(\eta^{\prime})\, d\eta^{\prime}],
\end{eqnarray}
where $\widehat{T}$ denotes time-ordering, is either $I$ or $Z$ in the odd $j$ outer intervals, 
\bes
\begin{eqnarray}
U_{c}(\eta) &=&I,\qquad\qquad [\eta_{j,2\ell },\eta_{j,2\ell +1}) \\
&=&Z,\qquad\qquad[\eta_{j,2\ell +1},\eta_{j,2\ell +2}),
\end{eqnarray}%
\ees
while in the even $j$ outer intervals,%
\bes
\begin{eqnarray}
U_{c}(\eta) &=&X,\qquad\qquad[\eta_{j,2\ell },\eta_{j,2\ell +1}) \\
&=&Y,\qquad\qquad[\eta_{j,2\ell +1},\eta_{j,2\ell +2}).
\end{eqnarray}
\ees

Accordingly, the normalized Hamiltonian in the toggling frame for the single-qubit general decoherence model,
\bes
\begin{eqnarray}
\widetilde{H}(\eta) &=&U_{c}(\eta)^{\dagger }HU_{c}(\eta) \\
&=&f_{0}J_{0}I\otimes B_{0}+f_{x}(\eta)J_{x}\sigma _{X}\otimes B_{X}\\
&&+f_{y}(\eta)J_{y}\sigma _{Y}\otimes B_{Y}+f_{z}(\eta)J_{z}\sigma _{Z}\otimes B_{Z} \notag,
\end{eqnarray}
\ees
has four different normalized QDD$_{N_{1} N_{2}}$ modulation functions,
\bes
\begin{eqnarray}
f_{0}&=&1 \qquad \qquad \qquad \qquad\quad  [0,1,), \label{f0}\\
f_{z}(\eta)&=&(-1)^{j-1}\qquad \qquad\quad \quad [\eta_{j-1},\eta_{j}),  \label{fz}\\
f_{x}(\eta)&=&(-1)^{k-1}\qquad \qquad\quad \quad [\eta_{j,k-1},\eta_{j,k}),  \label{fx}\\
f_{y}(\eta)&=&(-1)^{k-1}(-1)^{j-1}\qquad \,\,\,[\eta_{j,k-1},\eta_{j,k}),  \label{fy}\\
           &=&f_{x}(\eta)f_{z}(\eta)
\end{eqnarray}
\ees
unlike the single-qubit pure dephasing case, which has only two UDD modulation functions.  Because the $Z$-type pulses on the inner levels anti-commute with the errors $\sigma _{X}$ and $\sigma _{Y}$ and commute with $\sigma _{Z}$, the modulation functions $f_{x}(\eta)$ and $f_{y}(\eta)$ switch sign with the inner interval index $k$ while $f_{z}(\eta)$ is constant inside each outer interval. On the other hand, the outer $X$-type pulses anti-commute with the errors $\sigma _{Z}$ and $\sigma _{Y}$ and commute with the error $\sigma _{X}$, so both $f_{z}(\eta)$ and $f_{y}(\eta)$ switch sign with the outer interval index $j$,while $f_{x}(\eta)$ doesn't depend on the outer index $j$. 

Each QDD$_{N_{1} N_{2}}$ modulation function  $f_{\lambda}(\eta)$ can be separated naturally as 
\begin{equation}
f_{\lambda}(\eta)=f_{\tilde{\alpha}}(\eta)f_{\tilde{\beta}}(\eta)
\label{f in out}
\end{equation} 
where $f_{\tilde{\alpha}}(\eta)$  describes the behaviour of $f_{\lambda}(\eta)$ inside each outer interval  and $f_{\tilde{\beta}}(\eta)$ describes the behaviour of $f_{\lambda}(\eta)$ when the outer interval index $j$ changes. In fact $f_{\tilde{\beta}}(\eta)$ is identified as the normalized UDD$_{N_{2}}$ modulation function and $f_{\tilde{\alpha}}(\eta)$ covers $N_{2}+1$ cycles of UDD$_{N_{1}}$  modulation functions with different durations. However, up to a scale factor $f_{\tilde{\alpha}}(\eta)$ is the same function in each of these cycles. Therefore, instead of $f_{\tilde{\alpha}}(\eta)$,  we use one cycle of the normalized UDD$_{N_{1}}$  modulation function denoted as $f_{\alpha}(\eta)$ to denote the effective inner function of $f_{\lambda}(\eta)$. Likewise, since $f_{\tilde{\beta}}(\eta)$ is constant inside any $j$th outer interval $s_{j}$, it can be viewed as a function of the outer interval $j$, and we replace $f_{\tilde{\beta}}(\eta)$ by the notation $f_{\beta }(j)$. In particular, $f_{\beta=z}(j)=(-1)^{j-1}$.
Table~\ref{table:innerouter} lists the effective inner functions $f_{\alpha}$ and outer functions $f_{\beta}$ for all QDD$_{N_{1} N_{2}}$ modulation functions $f_{\lambda}$ and will be used in Sec. \ref{sec: ax and ay}.
\begin{table}[htbp]
\begin{tabular}{c|c}
\hline
$f_{\lambda}$ & $(f_{\alpha}, f_{\beta})$\\ \hline\hline
$f_{x}$ & $(f_{x}, f_{0})$\\ \hline
$f_{y}$ & $(f_{x}, f_{z})$\\ \hline
$f_{z}$ & $(f_{0}, f_{z})$\\ \hline
$f_{0}$ & $(f_{0}, f_{0})$\\ \hline
\end{tabular}
\caption{The effective inner functions $f_{\alpha}$ and outer functions $f_{\beta}$ of the normalized QDD$_{N_{1} N_{2}}$ modulation functions $f_{\lambda}$. Functions in the first column are the normalized QDD$_{N_{1} N_{2}}$ modulation function and those in the second column are the normalized  UDD$_{N_{1}}$ and  UDD$_{N_{2}}$ modulation functions respectively.}
\label{table:innerouter}\centering
\end{table}

\begin{table*}[htbp]
\begin{tabular}{|l|c|c|c|c|c|c|}
\hline
 \text{Components} \textbackslash \text{$n$th order QDD$_{N_{1} N_{2}}$ coefficients:} & \multicolumn{2}{|c|}{$a_{\lambda _{n}\cdots \lambda _{1}=X}$ } & \multicolumn{2}{|c|}{$a_{\lambda _{n}\cdots \lambda _{1}=Y}$ } & \multicolumn{2}{|c|}{$a_{\lambda _{n}\cdots \lambda _{1}=Z}$ }\\ \hline
\qquad \quad\;\,Total \# of $\sigma_{X}$ and $f_{x}(\eta)$  & odd & even & even & odd & even & odd \\ 
(1) \qquad Total \# of $\sigma_{Y}$ and $f_{y}(\eta)$  & even & odd & odd & even & even & odd \\ 
\qquad\quad \;\,Total \# of $\sigma_{Z}$ and $f_{z}(\eta)$  & even & odd & even & odd & odd & even \\ \hline\hline
$(2)$ Total \# of effective inner integrand $f_{x}$ & odd & odd & odd & odd & even & even \\ \hline
$(3)$ Total \# of effective outer integrand $f_{z}$ & even & even & odd & odd & odd & odd  \\ \hline
\end{tabular}
\caption{Number combinations of Pauli matrices (or modulation functions) for each error type (or QDD$_{N_{1} N_{2}}$ coefficients). For example, when ${\lambda _{n}\cdots \lambda _{1}=X}$, there are two possibilities, represented in the two corresponding columns in the rows numbered (1): either there is an odd number of $\sigma_X$ (and $f_x(\eta)$) along with an even number of both $\sigma_Y$ (and $f_y(\eta)$) and  $\sigma_Z$ (and $f_z(\eta)$) , or there is an even number of $\sigma_X$ (and $f_x(\eta)$) along with an odd number of both $\sigma_Y$ (and $f_y(\eta)$) and  $\sigma_Z$ (and $f_z(\eta)$) . Consulting Table~\ref{table:innerouter}, in the first case there is an odd number of inner integrand $f_x$ functions from $f_x(\eta)$ and an even number of $f_x$ from $f_y(\eta)$, so that the total number of $f_x$ is odd, as indicated in the first entry in row (2). Likewise, in the second case there is an odd number of outer integrand $f_z$ functions from $f_y(\eta)$ and an odd number of $f_z$ from $f_z(\eta)$, so that the total number of $f_z$ is even, as indicated in the second entry in row (3).}
\label{table:pauli}\centering
\end{table*}

In the toggling frame, the unitary evolution operator which contains a whole QDD$_{N_{1} N_{2}}$ sequence at the final time $T$ reads 
\bes
\bea
\widetilde{U}(T) &=& \widehat{T}\exp [-i\int_{0}^{T}\,\widetilde{H}(t)\,dt]  \\
&=& \widehat{T}\exp [-iT\int_{0}^{1}\,\widetilde{H}(\eta)\,d\eta].
\eea
\ees
We expand the evolution operator $\widetilde{U}(T)$ into the Dyson series of standard time dependent perturbation theory, 
\begin{equation}
\widetilde{U}(T) =\sum_{n=0}^{\infty }\sum_{\vec{\lambda}_n}(-iT)^{n}J_\lambda^{(n)}\sigma_\lambda^{(n)}\otimes B_\lambda^{(n)}a_{\lambda _{n}\cdots \lambda _{1}},  
\label{U}
\end{equation}
where we use the shorthand notation
\begin{equation}
\sum_{\vec{\lambda}_n}\equiv\sum_{\lambda _{n}\in\{0,X,Y,Z\}}\sum_{\lambda _{n-1}\in\{0,X,Y,Z\}}\dots \sum_{\lambda _{1}\in\{0,X,Y,Z\}}  ,
 \end{equation}
and
\begin{eqnarray}
J_\lambda^{(n)} \equiv \prod_{i=1}^n J_{\lambda_i}, \quad
\sigma_\lambda^{(n)} \equiv \prod_{i=1}^n \sigma_{\lambda_i}, \quad
B_\lambda^{(n)} \equiv \prod_{i=1}^n B_{\lambda_i}.
 \end{eqnarray}
Finally,
\begin{equation}
\ignore{
a_{\lambda _{n}\dots\lambda _{1}}\equiv \int_{0}^{1}d\eta^{(n)}\int_{0}^{\eta^{(n)}}d\eta ^{(n-1)}\dots\int_{0}^{\eta^{(2)}}d\eta^{(1)}\prod_{\ell=1}^{n}f_{\lambda _{\ell }}(\eta^{(\ell) }),  
}
a_{\lambda _{n}\cdots \lambda _{1}}\equiv \int_{0}^{1}d\eta^{(n)}
\dots\int_{0}^{\eta^{(2)}}d\eta^{(1)}\prod_{\ell=1}^{n}f_{\lambda _{\ell }}(\eta^{(\ell) })
\label{a}
\end{equation}
is a dimensionless constant we call the $n$th order normalized QDD$_{N_{1} N_{2}}$  coefficient. These coefficients play a key role in the theory as it is their vanishing which dictates the decoupling properties of the QDD sequence. 

The subscript of $a_{\lambda _{n}\cdots \lambda _{1}}$ represents a product of Pauli matrices, and we shall write $\lambda_{n}\cdots \lambda _{1} = \lambda$ with $\lambda$ representing the result of the multiplication up to $\pm 1, \pm i$.  From Eqs.~\eqref{U}-\eqref{a}, this subscript indicates not only its associated operator term  $\sigma_{\lambda_{n}}\dots\sigma_{\lambda_{1}}\otimes B_{\lambda_{n}}\cdots B_{\lambda _{1}}$ but also its $n$ ordered integrands, $f_{\lambda_{n}}\cdots f_{\lambda_{1}}$. Moreover, from Table \ref{table:innerouter}, one can also deduce the ordered set of effective inner and outer integrands for a given subscript of 
$a_{\lambda _{n}\cdots \lambda _{1}}$.


\subsection{Error terms}

Every product of system operators, $\sigma_\lambda^{(n)} = \sigma _{\lambda _{n}}\cdots \sigma _{\lambda _{1}}$ can be either $I$, $\sigma_{X}$, $\sigma_{Y}$ or $\sigma_{Z}$. The summands in the expansion (\ref{U}) of $\widetilde{U}$ can accordingly all be classified as belonging to one of four groups . If $\sigma_\lambda^{(n)} \in \{\sigma_X,\sigma_Y,\sigma_Z\}$ then the corresponding summand in Eq.~(\ref{U}) decoheres the system qubit. 

\begin{mydef}
A single-axis error of order $n$ and type $\lambda$ is the sum of all terms in Eq.~\eqref{U} with fixed $\sigma_\lambda^{(n)}$ and $\lambda\in \{X,Y,Z\}$.
\end{mydef}
In the UDD case Eq.~(\ref{U}) would include just one type of single-axis error \cite{YangLiu:08}.

Due to  the Pauli matrix identities $\sigma _{i}\sigma _{j}=i\varepsilon _{ijk}\sigma _{k}$ and $\sigma _{i}^{2}=I$, which of the three possible errors a given product $\sigma_\lambda^{(n)}$ becomes is uniquely determined by the parity of the total number of times each Pauli matrix appears in the product. In this sense there are only two possible ways in which each type of error can be generated. Take the error $\sigma _{X}$ as an example. One way to generate $\sigma _{X}$ is to have an odd number of $\sigma _{X}$ operators which generates $\sigma _{X}$ itself, along with an even number of $\sigma _{Y}$, an even number of $\sigma _{Z}$, and arbitrary number of $I$. The other possibility is an odd number of $\sigma_{Y}$ with an odd number of $\sigma _{Z}$ to generate $\sigma _{X}$, along with an even number of $\sigma _{X}$ and arbitrary number of $I$. 

Note that for a given error $\sigma _{\lambda}$, the parity of the total number of times each modulation function appears in $a_{\lambda _{n}\cdots \lambda _{1}=\lambda}$'s integrands $f_{\lambda _{n}}\cdots f_{\lambda _{1}}$ is also determined accordingly.  For example, consider ${\lambda _{n}\cdots \lambda _{1}=Z}$. This can be the result of there being an even number of $\sigma_Z$ [and $f_z(\eta)$] along with an odd number of both $\sigma_X$ [and $f_x(\eta)$] and  $\sigma_Y$ [and $f_y(\eta)$], a situation summarized in the last column of the block numbered (1) in Table~\ref{table:pauli}. In this case, given Table~\ref{table:innerouter}, the total number of effective inner integrand functions $f_x$ contributed by $f_x(\eta)$ is odd, as is the contribution of effective inner integrand functions $f_x$ from $f_y(\eta)$, so the total number of effective inner integrand functions $f_x$ is even. This situation is summarized by the last ``even'' entry in row (2) of Table~\ref{table:pauli}. This table gives all possible parities of Pauli matrices (or modulation functions) for each type of error (or its associated $n$th order QDD$_{N_{1} N_{2}}$  coefficient). The parity of the total number of identity matrices $I$ (modulation function $f_{0}$) is irrelevant for the proof, so is omitted from Table~\ref{table:pauli}. With a given number combination of Pauli matrices (or modulation functions) and Table \ref{table:innerouter}, one can determine the parity of the total number of effective inner and outer integrands as presented in rows (2) and (3) of Table \ref{table:pauli}. Table \ref{table:pauli} will be referred to often during the proof.

If all of the first $N$th order  QDD$_{N_{1} N_{2}}$  $\sigma_{\lambda}$ coefficients vanish for a given $\lambda$, namely $a_{\lambda _{n}\cdots \lambda _{1}=\lambda}=0$ for $n \leq N$,  we say that the QDD$_{N_{1} N_{2}}$ scheme eliminates the error $\sigma_{\lambda}$ to order $N$, i.e., the error $\sigma_{\lambda}$ is $\mathcal{O}(T^{N+1})$. Naively, one might expect the inner $Z$-type UDD$_{N_{1}}$ sequence to eliminate both $\sigma_X$ and $\sigma_Y$ errors to order $N_1$, and the outer $X$-type UDD$_{N_{2}}$ sequence to eliminate $\sigma_Z$ errors and any remaining $\sigma_Y$ errors to order $N_2$. The situation is in fact more subtle, and is summarized in the following QDD Theorem whose proof is provided in  Sec. \ref{sec: ax and ay} and Sec. \ref{sec:outer}.

\begin{QDD}
Assume that a single qubit is subject to the general decoherence model Eq.~\eqref{H}. Then, under the QDD$_{N_{1} N_{2}}$ sequence Eq.~\eqref{qddsequence}, 
all three types of single-axis errors of order $n$ are guaranteed to be eliminated if $n \leq \min[N_{1},N_{2}]$. Higher order single axis errors are also eliminated depending on the parities and relative magnitudes of $N_1$ and $N_2$, as detailed in Table \ref{table:qdd}, the results of which remain valid under any permutation of the labels $X,Y,Z$ along with a corresponding label permutation in Eq.~\eqref{qddsequence}.
\label{QDD}
\end{QDD}

\begin{table}[htbp]
\begin{tabular}{|c|c|c|l|}
\hline
Single-axis \ignore{error} & Inner order \ignore{$N_{1}$} & Outer order \ignore{$N_{2}$}  & Decoupling order  \\ 
error type& $N_1$ & $N_2$ &\\ \hline \hline
$\sigma_{X}$ & arbitrary & arbitrary & $N_{1}$ \\\hline 
\multirow{4}{*}{$\sigma_{Y}$} & even & even & $ \max[N_{1},N_2]$ \\  
& even & odd &  $ \max[N_{1}+1,N_2]$\\ 
& odd & even &  $ N_{1}$\\ 
& odd & odd &  $N_{1}+1$ \\ \hline
\multirow{2}{*}{$\sigma_{Z}$} & even & arbitrary &  $N_{2}$\\ 
& odd & arbitrary &   $\min[N_{1}+1,N_2]$ \\\hline
\end{tabular}
\caption{
Summary of single-axis error suppression. For each error type $\sigma_\lambda$, the $n$th order QDD$_{N_1 N_2}$ coefficients [Eq.~\red{\eqref{a}}] $a_{\lambda _{n}\cdots \lambda _{1}=\lambda}=0$  $\forall n\leq N$, where $N$ is the decoupling order given in the last column.}
\label{table:qdd}
\centering
\end{table} 

An immediate corollary of this Theorem is that the overall error suppression order of QDD$_{N_{1} N_{2}}$ is $\min[N_{1},N_{2}]$. This will be reflected in distance or fidelity measures for QDD, such as computed for UDD in Ref.~\cite{UL:10}.

We shall prove Theorem \ref{QDD} in two steps. First, in Sec.~\ref{sec: ax and ay} we shall prove that for arbitrary values of $N_1$ and $N_2$ the QDD$_{N_1 N_2}$ sequence eliminates the first $N_1$ orders of $\sx$ and $\sy$ errors. Secondly, we shall prove that if $N_2$ is odd, an additional order of the $\sy$ error is eliminated, i.e., $N_1+1$. We will not show any suppression of the $\sz$ error  in Sec.~\ref{sec: ax and ay}.

Then, in Sec.~\ref{sec:outer} we shall complete the analysis of the effect of the outer sequence, and show that the $\sz$ error is suppressed to order $N_2$ if $N_1$ is even. If $N_1$ is odd, $\sz$ is suppressed to order $N_2$ if $N_1 \geq N_2-1$, and to order $N_1+1$ if $N_1 < N_2-1$. Additionally, we show that if $N_1$ is even, the $\sy$ error is suppressed to order $N_2$, which may be higher than the result of Sec.~\ref{sec: ax and ay} alone. Combining the results of the two sections, we find that the error $\sy$ is suppressed to order $\max[N_1,N_2]$ if $N_2$ is even, and to order  $\max[N_1+1,N_2]$ if $N_2$ is odd. These results are all summarized in Table~\ref{table:qdd}.

\section{Suppression of longitudinal relaxation $\sigma _{X}$ and $\sigma _{Y}$}
\label{sec: ax and ay}

A general proof of the error suppression properties of UDD was first given by Yang \& Liu, including for the suppression of longitudinal relaxation errors $\sigma _{X}$ and $\sigma _{Y}$ \cite{YangLiu:08}. Wang \& Liu first proved that the outer sequence does not interfere with the suppression abilities of the inner sequence \red{with the DD pulses chosen as a MOOS set} \cite{WangLiu:11}.
In this section, we give an alternative non-interference proof which shows explicitly that it is the inner $Z$-type UDD$_{N_{1}}$ sequence that makes all longitudinal relaxation related QDD$_{N_{1} N_{2}}$ coefficients $a_{\lambda _{n}\cdots \lambda _{1}=\sigma _{X},\sigma _{Y}}$ with $n\leq N_1$ vanish, regardless of the details of the outer $X$-type UDD$_{N_{2}}$ sequence. Moreover, we also show that the outer $X$-type UDD$_{N_{2}}$ sequence, when the outer order $N_2$
is odd, eliminates the $\sigma _{Y}$ error to one additional order, i.e., to order $N_1+1$. For precise details refer to Table~\ref{table:qdd}. 

\subsection{The outer interval decomposition of $a_{\lambda _{n}\cdots \lambda _{1}}$}

We expect the inner $Z$-type UDD$_{N_{1}}$ sequences of QDD$_{N_{1} N_{2}}$ to suppress the errors $\sigma _{X}$ and $\sigma _{Y}$. Therefore, our  strategy for evaluating $a_{\lambda _{n}\cdots \lambda _{1}}$ [Eq.~\eqref{a}] is to split each of its integrals into a sum of sub-integrals over the normalized outer intervals $s_{j}$ in Eq.~\eqref{sj}. In this way,  each resulting segment of $a_{\lambda _{n}\cdots \lambda _{1}}$ can be decomposed naturally into an inner part (which contains the action of the inner $Z$-type UDD$_{N_{1}}$) times an outer part (which contains the action of the outer $X$-type UDD$_{N_{2}}$ sequence).  The manner by which the inner $Z$-type UDD$_{N_{1}}$ sequences suppress longitudinal relaxations can then be easily extracted. 

As we show in Appendix \ref{app:form a}, 
after this decomposition $a_{\lambda _{n}\cdots \lambda _{1}}$ can be expressed as  
\begin{eqnarray}
&&a_{\lambda _{n}\cdots \lambda _{1}}=\sum_{\{r_{\ell}=\emptyset,\ast\}{_{\ell=1}^{n-1}}}\Phi^{\rm in}( r_{n} f_{\alpha _{n}}r_{n-1}\dots f_{\alpha _{2}} r_{1} f_{\alpha _{1}})\times \notag \\
&&\quad \Phi^{\rm out}( r_{n} f_{\beta _{n}}r_{n-1}\dots f_{\beta _{2}} r_{1} f_{\beta _{1}})
\label{form a}.
\end{eqnarray}
with $r_{n}\equiv\ast$. This is just a compact way of writing multiple nested integrals and multiple summations, with a notation we explain next.

First, $f_{\alpha _{\ell}}$ and $f_{\beta _{\ell}}$ are the effective inner and outer functions respectively of $a_{\lambda_{n}\cdots \lambda_{1}}$'s $\ell$th  integrand $f_{\lambda_{\ell}}$. From Table \ref{table:innerouter}, the effective inner (outer) function of the normalized QDD$_{N_{1} N_{2}}$ modulation functions will be either $f_{x}$  ($f_{z}$) or $f_{0}=1$, the normalized  UDD$_{N_{1}}$ (UDD$_{N_{2}}$) modulations functions in the generic $\sigma_{X}$ ($\sigma_{Z}$) pure bit flip (dephasing) model. 

Second, the ``inner output function'' $\Phi^{\rm in}$  generates  all the segments' inner parts via the following mapping, 
\begin{eqnarray}
r_{\ell}f_{\alpha _{\ell}} &\xmapsto{\Phi^{\rm in}}&
             \begin{cases}
             \int_{0}^{1}f_{\alpha _{\ell}}(\eta^{(\ell)})\,d\eta^{(\ell)}                & \text{if } r_{\ell}=\ast \\
              \int_{0}^{\eta^{(\ell+1)}}f_{\alpha _{\ell}}(\eta^{(\ell)})\,d\eta^{(\ell)} & \text{if } r_{\ell}=\emptyset.
             \end{cases}
             \label{eq:inner}
\end{eqnarray}
For example, 
\begin{equation}
\Phi^{\rm in}(\ast f_{\alpha_2} \emptyset f_{\alpha_1}) = \int_0^1 f_{\alpha_2}d\eta^{(2)}\int_0^{\eta^{(2)}} f_{\alpha_1}d\eta^{(1)} ,
\end{equation}
a term which appears in the expansion of $a_{\lambda_2 \lambda_1}$.

From Eq.~(\ref{eq:inner}), one can see that $r_{\ell}$ determines how the integral of $\eta^{(\ell)}$ relates to the integral of its adjacent variable  $\eta^{(\ell+1)}$. For the inner part, the relationship between the integrals of two adjacent variables $\eta^{(\ell+1)}$ and  $\eta^{(\ell)}$ is either independent ($r_{\ell}=\ast$; they appear in separate integrals) or nested ($r_{\ell}=\emptyset$; they appear together in a time-ordered pair of integrals).

Third, the outer output function $\Phi^{\rm out}$  generates  all the segments' outer parts via the following mapping, 
\begin{eqnarray}
r_{\ell}f_{\beta _{\ell}} &\xmapsto{\Phi^{\rm out}}&
             \begin{cases}
            \sum_{j^{(\ell)}=m}^{j^{(\ell+1)}-1} f_{\beta _{\ell}}(j^{(\ell)})\,s_{j^{(\ell)}}             & \text{if } r_{\ell}=\ast \\
              f_{\beta_{\ell}}(j^{(\ell+1)})\,s_{j^{(\ell+1)}}                                              & \text{if } r_{\ell}=\emptyset
             \end{cases}                                 
\label{eq:outer}
\end{eqnarray}
where $s_{j^{(\ell)}}$ is the $j^{(\ell)}$th normalized outer interval for variable $\eta^{(\ell)}$, and $m$ indicates that $r_{\ell}$ is the $m$th $\ast$ in $\{r_{n}r_{n-1}\dots r_{1}\}$, counting from $r_{1}$. For $r_{n}f_{\beta _{n}}$ with $r_{n}=\ast$, the upper limit $j^{(\ell)}=j^{(\ell+1)}-1$ in Eq.~\eqref{eq:outer} should be replaced by $j^{(n)}=N_{2}+1$. 
For example, 
\begin{eqnarray}
&&\Phi^{\rm out}(\ast f_{\beta_3} \ast f_{\beta_2} \emptyset f_{\beta_1}) = \sum_{j^{(3)}=2}^{N_2+1} f_{\beta_3}(j^{(3)})s_{j^{(3)}} \times \notag \\
&&\qquad\sum_{j^{(2)}=1}^{j^{(3)}-1} f_{\beta_2}(j^{(2)})s_{j^{(2)}} 
f_{\beta_1}(j^{(2)})s_{j^{(2)}},
\end{eqnarray}
a term which appears in the expansion of $a_{\lambda_3 \lambda_2 \lambda_1}$.

From Eq.~\eqref{eq:outer}, $r_{\ell}$ indicates the relationship between the outer intervals of two adjacent variables $\eta^{(\ell+1)}$ and $\eta^{(\ell)}$. They can either be time-ordered, namely, in different outer intervals ($r_{\ell}=\ast$), or in the same interval ($r_{\ell}=\emptyset$).

Finally, 
\begin{equation}
\sum_{\{r_{\ell}=\emptyset,\ast\}{_{\ell=1}^{n-1}}}\equiv \sum_{r_{n-1}\in\{\emptyset,\ast\}}\sum_{r_{n-2}\in\{\emptyset,\ast\}}\dots\sum_{r_{1}\in\{\emptyset,\ast\}}
\end{equation}
includes all possible integration configurations for $\Phi^{\rm in}$ and all possible summation configurations for $\Phi^{\rm out }$. Each such configuration is determined by a given set $\{\ast,r_{n-1},r_{n-2},\dots, r_{1}\}$.

Note that the integration pattern of the inner part determines the summation pattern of the outer part and vice versa. The relation between the inner part and its corresponding outer part comes from the time-ordering condition, $\eta^{(n)}\geq\eta^{(n-1)}\geq \dots \eta^{(2)}\geq\eta^{(1)}$, because $a_{\lambda _{n}\cdots \lambda _{1}}$ comprises $n$ time-ordered integrals.  More specifically, if the sub-integrals over any two adjacent variables $\eta^{(\ell)}$ and $\eta^{(\ell+1)}$ are already located in time-ordered, different outer intervals, then the sub-integral over $\eta^{(\ell)}$ is not nested inside the sub-integral over $\eta^{(\ell+1)}$, and its integration domain is the entire outer interval. In contrast,
if the sub-integrals over any two adjacent variables $\eta^{(\ell)}$ and $\eta^{(\ell+1)}$ are in the same outer interval, it follows that their sub-integrals are nested due to time-ordering.


\subsection{The inner parts $\Phi^{\rm in}$ and the outer parts $\Phi^{\rm out}$ of $a_{\lambda _{n}\cdots \lambda _{1}}$}

Consider the argument $r_{n} f_{\mu_{n}}r_{n-1}f_{\mu _{n-1}}r_{n-2}\dots f_{\mu_{2}} r_{1} f_{\mu_{1}}$ of $\Phi^{\rm in}$ or $\Phi^{\rm out}$, where $\mu$ can be $\alpha$ or  $\beta$ in Eq.~\eqref{form a}. Define a cluster of $f$'s as a contiguous set of $f$'s connected only by $\emptyset$. Different clusters are separated by $\ast$. For example, $(f_{\mu_5})\ast (f_{\mu_4}\emptyset f_{\mu_3})\ast (f_{\mu_2}\emptyset f_{\mu_1})$, where the parentheses indicate clusters. 
\ignore{
In the next step we shall discard the $\emptyset$ symbols and keep only the $\ast$. Thus, in the example above we have $(f_{\mu_5})\ast (f_{\mu_4} f_{\mu_3})\ast (f_{\mu_2} f_{\mu_1})$.
}
In this manner, each integration or summation configuration $\{\ast r_{n-1}r_{n-2}\dots r_{1}\}$  corresponds to a way in which a set of $n$ functions is separated into clusters. 

Suppose that for a given configuration  $\{\ast r_{n-1}r_{n-2}\dots r_{1}\}$, the $m$th inner cluster, counting from right to left, is $f_{\alpha_{p}}\emptyset f_{\alpha_{p-1}}\emptyset \dots \emptyset f_{\alpha_{q}}$, which has  $p-q+1$ elements.  Likewise, we have the $m$th outer cluster  $f_{\beta_{p}}\emptyset f_{\beta_{p-1}}\emptyset\dots\emptyset f_{\beta_{q}}$. 
Applying the rule of Eq.~\eqref{eq:inner} to the $m$th inner cluster, or the rule of Eq.~\eqref{eq:outer} to the $m$th outer cluster, we then have, respectively
\bes
\begin{eqnarray}
&\ast& f_{\alpha_{p}}\emptyset \dots \emptyset f_{\alpha_{q}} (\ast) \xmapsto{\Phi^{\rm in}} \label{mapgroup1} \\
&&  \int_{0}^{1}d\eta^{(p)} \int_{0}^{\eta^{(p)}}d\eta^{(p-1)}\dots\int_{0}^{\eta^{(q+1)}}d\eta^{(q)}\prod_{\ell=q}^{p}f_{\alpha_{\ell }}(\eta^{(\ell) }) \notag \\
&& \equiv u_{\alpha_{p}\alpha_{p-1}\dots\alpha_{q}} \\
&\ast& f_{\beta_{p}}\emptyset  \dots \emptyset f_{\beta_{q}}(\ast)  \xmapsto{\Phi^{\rm out}}  \sum_{j_{m}=m}^{j_{m+1}-1} \prod_{\ell=q}^{p}f_{\beta _{\ell}}(j_{m})\,s_{j_{m}}^{p-q+1}
\label{mapgroup}
\end{eqnarray}
\ees
where if $p=n$, namely the $m$th group is the last group (counting from right to left), the upper limit $j_{m+1}-1$ should be replaced by $j_{m}=N_{2}+1$. Also note that in Eq.~\eqref{mapgroup} we have replaced $j^{(p)}$ [according to the notation of Eq.~\eqref{eq:outer}] by the cluster index $j_m$.

Now recall that the outer effective function $f_{\beta _{\ell}}(j)$ is either $f_{0}=1$ or $f_{z}(j)=(-1)^{j-1}$. Therefore, if $\prod_{\ell=q}^{p}f_{\beta _{\ell}}(j_{m})$ contains an odd number of $f_{z}(j_m)$, we have $\prod_{\ell=q}^{p}f_{\beta _{\ell}}(j_{m})=(-1)^{j_{m}-1}$, otherwise $\prod_{\ell=q}^{p}f_{\beta _{\ell}}(j_{m})=1$. 

Note that the nested integral $u_{\alpha_{p}\alpha_{p-1}\dots\alpha_{q}}$ in Eq.~\eqref{mapgroup1} is just the $(p-q+1)$th order normalized UDD$_{N_{1}}$ coefficient for the generic $\sigma_{X}$ pure bit flip model, because the effective integrands $f_{\alpha_{\ell}}$, either $f_{x}$ or $f_{0}$, are the normalized UDD$_{N_{1}}$  modulations functions. 

We have now assembled the tools to perform the summation implied in Eq.~\eqref{form a}, which is the result of the outer interval decomposition. Different clusters, each of which is given in Eqs.~\eqref{mapgroup1} or \eqref{mapgroup}, are simply multiplied. To illustrate this, the second order normalized QDD$_{N_{1} N_{2}}$ coefficients $a_{\lambda _{1}\lambda _{2}}$ are listed in Table \ref{table:aorder2}.
\begin{table*}[htbp]
\begin{tabular}{|c|l|}
\hline
  Error Type & 2nd order normalized QDD$_{N_{1} N_{2}}$ coefficients\\ \hline
\multirow{4}{*}{$\sigma_{X}$} &
  $a_{x0}=u_{x}\,u_{0}\sum_{j=2}^{N_{2}+1}s_{j} \,\sum_{p=1}^{p<j} s_{p} + u_{x0}\sum_{j=1}^{N_{2}+1}s_{j}^{2}$ \\
  & $ a_{0x}=u_{0}\,u_{x}\sum_{j=2}^{N_{2}+1}s_{j} \,\sum_{p=1}^{p<j} s_{p} +  u_{0x}\sum_{j=1}^{N_{2}+1}s_{j}^{2}$\\ 
  & $a_{zy}=u_{0}\,u_{x}\sum_{j=2}^{N_{2}+1}(-1)^{j-1} s_{j}  \,\sum_{p=1}^{p<j}(-1)^{p-1} s_{p} +u_{0x}\sum_{j=1}^{N_{2}+1}s_{j}^{2}$\\ 
  & $a_{yz}=u_{x}\,u_{0}\sum_{j=2}^{N_{2}+1}(-1)^{j-1} s_{j} \,\sum_{p=1}^{p<j}(-1)^{p-1} s_{p} + u_{x0}\sum_{j=1}^{N_{2}+1}s_{j}^{2}$  \\ \hline
 \multirow{4}{*}{$\sigma_{Y}$ } & $a_{y0}=u_{x}\,u_{0}\sum_{j=2}^{N_{2}+1}(-1)^{j-1}s_{j}\,\sum_{p=1}^{p<j} s_{p} + u_{x0}\sum_{j=1}^{N_{2}+1}(-1)^{j-1}s_{j}^{2}$ \\ 
   & $ a_{0y}=u_{0}\,u_{x}\sum_{j=2}^{N_{2}+1}s_{j} \,\sum_{p=1}^{p<j} (-1)^{p-1}s_{p} +u_{0x}\sum_{j=1}^{N_{2}+1}(-1)^{j-1}s_{j}^{2}$\\ 
   & $a_{zx}=u_{0}\,u_{x}\sum_{j=2}^{N_{2}+1}(-1)^{j-1} s_{j} \,\sum_{p=1}^{p<j} s_{p} +u_{0x}\sum_{j=1}^{N_{2}+1}(-1)^{j-1}s_{j}^{2}$\\ 
   & $a_{xz}=u_{x}\,u_{0}\sum_{j=2}^{N_{2}+1}s_{j} \,\sum_{p=1}^{p<j}(-1)^{p-1} s_{p} +u_{x0}\sum_{j=1}^{N_{2}+1}(-1)^{j-1}s_{j}^{2}$  \\ \hline
  \multirow{4}{*}{$\sigma_{Z}$} & $a_{z0}=u_{0}\,u_{0}\sum_{j=2}^{N_{2}+1}(-1)^{j-1}s_{j} \,\sum_{p=1}^{p<j} s_{p} + u_{00}\sum_{j=1}^{N_{2}+1}(-1)^{j-1}s_{j}^{2}$ \\ 
  & $ a_{0z}=u_{0}\,u_{0}\sum_{j=2}^{N_{2}+1}s_{j} \,\sum_{p=1}^{p<j}(-1)^{p-1} s_{p} +  u_{00}\sum_{j=1}^{N_{2}+1}(-1)^{j-1}s_{j}^{2}$\\ 
 & $a_{xy}=u_{x}\,u_{x}\sum_{j=2}^{N_{2}+1}s_{j}  \,\sum_{p=1}^{p<j}(-1)^{p-1} s_{p} +u_{xx}\sum_{j=1}^{N_{2}+1}(-1)^{j-1}s_{j}^{2}$\\ 
   & $a_{yx}=u_{x}\,u_{x}\sum_{j=2}^{N_{2}+1}(-1)^{j-1} s_{j} \,\sum_{p=1}^{p<j} s_{p} + u_{xx}\sum_{j=1}^{N_{2}+1}(-1)^{j-1}s_{j}^{2}$  \\ \hline
\end{tabular}
\caption{The outer decomposition form of $a_{\lambda _{1}\lambda _{2}}$. For $n=2$ we have $r_2 r_1 = \{\ast \ast\}$ or $r_2 r_1 = \{\ast \emptyset\}$. The first summand in each line is the result of the $\{\ast \ast\}$ expansion, the second is the result of the $\{\ast \emptyset\}$ expansion.} 
\label{table:aorder2}
\end{table*}

The following lemmas relate the normalized QDD and UDD coefficients. They are easily concluded from Eq.~\eqref{mapgroup1}. 

Consider a configuration $\{\ast r_{n-1}r_{n-2}\dots r_{1}\}$ with $m$ $\ast$'s. Correspondingly there are $m$ clusters. Each cluster has associated with it a normalized UDD$_{N_{1}}$ coefficient of order $n'$ equal to the number of elements ($f$'s) in the cluster, and $1\leq n' <n$.  The sum of all the orders is $n$.  Thus:
\begin{mylemma}
Consider a configuration $\{\ast r_{n-1}r_{n-2}\dots r_{1}\}$ with $m$ $\ast$'s. The corresponding inner part $\Phi^{\rm in}$ of
$a_{\lambda _{n}\cdots \lambda _{1}}$ [Eq.~\eqref{form a}] is composed of $m$  normalized UDD$_{N_{1}}$ coefficients whose integrands are the effective inner ones of $a_{\lambda _{n}\cdots \lambda _{1}}$, and  the sum of whose orders is equal to $n$.
\label{lem:inner a udd}
\end{mylemma}

In addition, all of the first $n$th order UDD$_{N_1}$ coefficients, but not order $n+1$ and above, appear in any given $n$th order QDD$_{N_1 N_2}$ coefficient, i.e.,
\begin{mylemma} 
The only UDD$_{N_{1}}$ coefficients which can appear in all the inner parts $\Phi^{\rm in}$ of the $n$th order QDD$_{N_{1} N_{2}}$ coefficient $a_{\lambda _{n}\cdots \lambda _{1}}$ are those whose orders are between $1$ and $n$.
\label{lem:inners a udd}
\end{mylemma}


\subsection{The first $N_{1}$ vanishing orders  of the single-axis $\sigma_X$ and $\sigma_Y$ error due to the inner $Z$-type UDD$_{N_{1}}$ sequences}

From the second row of Table \ref{table:pauli}, one can see that the total number of $f_{x}$'s in the effective inner integrands of  the coefficients $a_{\lambda _{n}\cdots \lambda _{1}=X}$  and $a_{\lambda _{n}\cdots \lambda _{1}=Y}$ is odd. Accordingly, no matter how one divides the inner integrands into clusters, there will always be at least one cluster which has an odd number of $f_{x}$.  Then, from Lemma \ref{lem:inner a udd}, it follows that all the inner parts $\Phi^{\rm in}$ of $a_{\lambda _{n}\cdots \lambda _{1}=X}$  and $a_{\lambda _{n}\cdots \lambda _{1}=Y}$ contain one or more UDD$_{N_{1}}$ coefficients with an odd number of $f_{x}$ in the integrands.   Recall that UDD$_{N_{1}}$ coefficients $u_{\lambda _{m}\dots\lambda _{1}=X}$, i.e., those associated with the error $\sigma_{X}$, contain an odd number of $f_{x}$ in their integrands. Therefore, we have

\begin{mylemma}
After outer interval decomposition, all the inner parts of the $n$th order QDD$_{N_{1} N_{2}}$ coefficients $a_{\lambda_{n}\dots\lambda _{1}=X}$  and $a_{\lambda _{n}\dots\lambda _{1}=Y}$ contain one or more UDD$_{N_{1}}$ coefficients $u_{\lambda _{m}\dots\lambda _{1}=X}$, where  $m \leq n $. 
\label{lem:axy}
\end{mylemma}

Now recall:
\begin{mylemma} (Yang \& Liu \cite{YangLiu:08})
The UDD$_{N_{1}}$ coefficients $u_{\lambda _{m}\dots\lambda _{1}=X}=0$ when $m \leq N_{1}$.
\label{lem:udd}
\end{mylemma}
It follows from the last two lemmas that all QDD$_{N_{1} N_{2}}$ coefficients associated with longitudinal relaxation $a_{\lambda _{n}\cdots \lambda _{1}=X}$ or $a_{\lambda _{n}\cdots \lambda _{1}=Y}$ with $n\leq N_{1}$ vanish. Physically, it is clearly the inner $Z$-type  UDD$_{N_{1}}$ sequence that is responsible for eliminating the single-axis errors $\sigma _{X}\otimes B_{X}$ and $\sigma _{Y}\otimes B_{Y}$ up to order $T^{N_{1}+1}$. The effect of the outer $X$-type UDD$_{N_{2}}$ sequence  is entirely contained in the outer output function $\Phi^{\rm out}$ in Eq.~\eqref{form a}, and consequently does not interfere with the elimination ability of the inner level control $Z$-type UDD$_{N_{1}}$, in agreement with \cite{WangLiu:11}.

From row 2 of Table \ref{table:pauli}, unlike  $a_{\lambda_{n}\cdots \lambda_{1}=X,Y}$,  all $a_{\lambda_{n}\cdots \lambda _{1}=Z}$ contain an even number of effective inner functions $f_{x}$. Accordingly, Lemma  \ref{lem:axy} does not apply to $a_{\lambda _{n}\cdots \lambda _{1}=Z}$ and therefore, the argument that all the inner output functions $\Phi^{\rm in}$ of $a_{\lambda_{n}\cdots \lambda_{1}}$ are removed by the $Z$-type UDD$_{N_{1}}$ sequences  cannot be applied to the pure dephasing terms. This is, of course, due to the fact that the inner $Z$-type sequence commutes with the the pure dephasing error $\sigma _{Z}\otimes B_{Z}$. Instead, this error will be suppressed  by the outer $X$-type UDD$_{N_{2}}$ sequence. 

Note that the outer output functions $\Phi^{\rm out}$ of  $a_{\lambda _{n}\cdots \lambda _{1}}$ [Eq.~\eqref{form a}], which contain the effect of the outer $X$-type UDD$_{N_{2}}$ sequence, are expressed in terms of multiple time-ordered summations [Eq.~\eqref{mapgroup}], which are not easily analyzed using the current method. Therefore, in order to demonstrate the suppression of the pure dephasing error $\sigma _{Z}\otimes B_{Z}$, in Sec.~\ref{sec:outer} we shall deal directly with $a_{\lambda _{n}\dots\lambda _{1}=Z}$, rather than a separation into inner and outer parts as we have done in this section.


\subsection{One more order of suppression for the single-axis $\sigma_Y$ error due to the outer $X$-type UDD$_{N_2}$ sequence when $N_2$ is odd}

In the previous subsection we proved that $a_{\lambda _{n}\cdots \lambda _{1}=Y}=0$ when $n \leq N_{1}$. Now we shall show that also $a_{\lambda _{N_{1}+1}\dots\lambda _{1}=Y}$ vanishes, due to the outer level $X$-type UDD$_{N_{2}}$ sequence, for  odd $N_{2}$. Essentially, as we now explain in detail, this sequence is responsible for eliminating one remaining term in the expansion of  $a_{\lambda _{N_{1}+1}\dots\lambda _{1}=Y}$.

According to Lemma~\ref{lem:inners a udd}, as applied to $a_{\lambda _{N_{1}+1}\dots\lambda _{1}=Y}$, the only UDD$_{N_1}$ coefficients which can appear are those with order at most $N_1+1$. According to Lemma~\ref{lem:udd} the first $N_1$ of these UDD coefficients vanish.  The only UDD coefficient (in $a_{\lambda _{N_{1}+1}\dots\lambda _{1}=Y}$) regarding which at this point we have no information is the $N_1+1$th, and indeed, it may be nonvanishing. Using the mapping Eq.~\eqref{mapgroup}, we therefore have
\begin{equation}
a_{\lambda _{N_{1}+1}\dots\lambda _{1}=Y}=u_{\lambda _{N_{1}+1}\dots\lambda _{1}=X}\sum_{j=1}^{N_{2}+1} \prod_{\ell=1}^{N_{1}+1}f_{\beta _{\ell}}(j)\,s_{j}^{N_{1}+1}.
\label{ayN+1}
\end{equation}
We now show that this vanishes due to the outer part.

According to the third row of  Table \ref{table:pauli}, all $a_{\lambda _{n}\cdots \lambda _{1}=Y}$ contain an odd number of effective outer functions $f_{z}(j)=(-1)^{j-1}$. Consequently, we have $\prod_{\ell=1}^{N_{1}+1}f_{\beta _{\ell}}(j)=(-1)^{j-1}$ which simplifies the outer part in Eq.~\eqref{ayN+1} to
\begin{equation}
\sum_{j=1}^{N_{2}+1} (-1)^{j-1}s_{j}^{N_{1}+1}.
\label{ayN1+1}
\end{equation}
Note that the UDD pulse intervals  are time-symmetric (for the proof see Appendix \ref{app:timesym}). Therefore, the UDD$_{N_{2}}$ outer intervals $s_{j}$ satisfy
\begin{equation}
s_{j}=s_{N_{2}+2-j}
\label{sym}
\end{equation}
If the decoupling order of the outer UDD sequence $N_{2}$ is odd then  $j$ and  $N_{2}+2-j$ have opposite parities. Accordingly, 
\begin{eqnarray}
(-1)^{j-1}s_{j}^{N_{1}+1}&=&(-1)^{j-1}(s_{N_{2}+2-j})^{N_{1}+1} \label{odd-cancel} \\
                          &=&-(-1)^{N_{2}+2-j-1}(s_{N_{2}+2-j})^{N_{1}+1}. \notag
\end{eqnarray}
Thus, when $N_2$ is odd the outer part Eq.~\eqref{ayN1+1} vanishes due to the mutual cancellation of terms with equal magnitude but opposite sign.

This concludes our proof of the error suppression of $\sigma_X$ and $\sigma_Y$ errors to order $N_1$, and of $\sigma_Y$ to order $N_1+1$ when $N_2$ is odd. This confirms row one of Table~\ref{table:qdd} and row two of the same Table, disregarding for now $N_2$ in the last column. In the next Section we set out to complete the proof and confirm all claims made in Table~\ref{table:qdd}.


\section{Suppression of the pure dephasing error $\sigma _{Z}$}
\label{sec:outer}

In this section we focus on the suppression of the pure dephasing error $\sigma _{Z}$ by the outer $X$-type sequence, and also show that $\sigma _{Y}$ can be additionally suppressed by the outer sequence to order $N_2$ when $N_1$ is even.

To do so, we shall show that if $N_1$ is even the inner $Z$-type UDD$_{N_{1}}$ sequence does not hinder the suppression ability of the $Y$ and $Z$-type errors by the outer $X$-type UDD$_{N_{2}}$ sequence. For odd $N_1$ we cannot conclude that the inner sequence does not hinder the outer sequence. However, if $N_1$ is odd, our method does show that the outer sequence suppresses $\sigma _{Z}$ at least to order $\min[N_{1}+1,N_{2}]$.


\subsection{Linear change of variables}

To avoid having to analytically integrate a multiple nested integral with step functions as integrands such as $a_{\lambda _{n}\cdots \lambda _{1}}$,  we adapt the approach of Refs.~\cite{YangLiu:08,WangLiu:11}, which avoids integrating step functions directly but still manages to show $a_{\lambda_{n}\cdots \lambda _{1}=Z}=0$ up to a certain order. 

First, we make an appropriate variable transformation from $\eta \in [0,1)$ to $\theta \in [0,\pi)$, with the result that the outer pulse intervals are all equal. This is required to make the modulation functions $f_{x}$, $f_{y}$, $f_{z}$, and $f_{0}$ (possible integrands that can occur in $a_{\lambda _{n}\cdots \lambda _{1}}$) become periodic functions so that each of their Fourier expansions is either a Fourier sine or Fourier cosine series.

The variable transformation introduced by \cite{WangLiu:11} to tackle the QDD$_{N_{1} N_{2}}$ sequence is to apply the corresponding  linear transformation  to each outer pulse interval $\lbrack \eta_{j-1},\eta_{j})$ with duration $s_{j}$, 
\begin{equation}
\theta =\frac{\pi }{N_{2}+1}\left(\frac{\eta-\eta_{j-1}}{s_{j}}\right)+\frac{(j-1)\pi}{N_{2}+1}.
\label{linear}
\end{equation}

The timing of the outer $X$-type pulses becomes 
\begin{equation}
\theta _{j}= \frac{j\pi }{N_{2}+1}
\end{equation} 
so that $f_{z}(\theta )$ becomes a periodic function with period of $\frac{2\pi }{^{N_{2}+1}}$,
\begin{eqnarray}
f_{z}(\theta ) &=&(-1)^{j-1} \qquad \qquad \qquad \qquad [\theta_{j-1},\theta _{j}) \notag \\
               &=&\sum_{k=0}^{\infty }d_k^z \sin [(2k+1)(N_{2}+1)\theta ],
\label{afz}
\end{eqnarray}
where the second equality is the Fourier sine-series expansion, and $d_k^z = \frac{4}{(2k+1)\pi}$.

When we apply the piecewise linear transformation \eqref{linear} to the inner pulse timings $\eta_{j,k}$ [Eq.~\eqref{eta-jk}] the UDD$_{N_{1}}$ structure is preserved
\begin{equation}
\theta _{j,k}=\frac{\pi }{N_{2}+1}\sin ^{2}\left(\frac{k\pi }{2(N_{1}+1)}\right)+\theta _{j-1}.
\label{thetajk}
\end{equation} In fact all the inner pulse sequences become identical as they have the same total duration $\frac{\pi }{^{N_{2}+1}}$. It follows that $f_{x}(\theta )=(-1)^{k-1}$  within $[\theta_{j,k-1},\theta _{j,k})$ is a periodic function with  period of $\frac{\pi }{^{N_{2}+1}}$.

The parity of the decoupling order $N_{1}$ of the inner UDD$_{N_{1}}$ sequence determines whether $f_{x}(\theta )$ is even or odd inside each outer interval (Appendix \ref{app:sys fx}). Inside each outer interval the parity of $f_{x}(\theta )$ equals that of $N_{1}$. Hence, we have
\begin{eqnarray}
f_{x}(\theta )=\begin{dcases*}
               \sum_{k=0}^{\infty }d_{k}^{x}\cos[2k(N_{2}+1)\theta ] &$N_{1}$ even\\
               \sum_{k=1}^{\infty }d_{k}^{x}\sin [2k(N_{2}+1)\theta ] &$N_{1}$ odd
               \end{dcases*}
\label{afx}
\end{eqnarray}%

The relation $f_{y}(\theta ) =f_{z}(\theta )f_{x}(\theta )$, Eqs.~\eqref{afz} and \eqref{afx}, and the product-to-sum rules of the trigonometric functions,
\begin{subequations}
\begin{eqnarray}
\sin a\sin b &=&\frac{1}{2}\left[ \cos \left( a-b\right) -\cos \left(
a+b\right) \right] ,  \label{sinsin} \\
\cos a\sin b &=&\frac{1}{2}\left[ \sin \left( a+b\right) -\sin \left(
a-b\right) \right] ,  \label{cossin} \\
\cos a\cos b &=&\frac{1}{2}\left[ \cos \left( a+b\right) +\cos \left(
a-b\right) \right] , \label{coscos}
\end{eqnarray}
\end{subequations}
yield the following Fourier expansions of $f_{y}(\theta )$
\begin{eqnarray}
f_{y}(\theta )=\begin{dcases*}
              \sum_{k=0}^{\infty }d_{k}^{y}\sin [(2k+1)(N_{2}+1)\theta ]& $N_{1}$ even\\
              \sum_{k=0}^{\infty }d_{k}^{y}\cos [(2k+1)(N_{2}+1)\theta ] & $N_{1}$ odd
               \end{dcases*}
               \label{afy}
\end{eqnarray}%

Note that while the Fourier expansion coefficients in the even and odd cases are in fact different, we use the notation $d_{k}^{x}$ or $d_{k}^{y}$ for both since the exact values of these coefficients are irrelevant for our proof.

It follows from Eq.~\eqref{linear} that  $d\eta =\frac{N_{2}+1}{\pi }s_{j}\,d\theta = G(\theta )d\theta$, where $G(\theta )$ is the step function whose step heights are proportional to the QDD$_{N_{1} N_{2}}$ outer intervals,
\bes
\begin{eqnarray}
\!\!\!\!\! G(\theta ) &=& \frac{N_{2}+1}{\pi }s_{j}  \qquad\qquad \qquad \theta \in \lbrack \theta _{j-1},\theta _{j}) \\
&=&\sum_{k=0}^{\infty }\sum_{q=-1,1}g_{k,q}\sin [(2k)(N_{2}+1)\theta +q\theta ],  \label{G}
\end{eqnarray}
\ees
as shown in Appendix \ref{app:G}.

With Eqs.~\eqref{afz}, \eqref{afx}-\eqref{G}, and  $f_{0}=1$, the $n$th order QDD$_{N_{1} N_{2}}$ coefficients \eqref{a} can be rewritten as
\begin{equation}
a_{\lambda _{n}\cdots \lambda _{1}}= \int_{0}^{\pi }d\theta_{n}\int_{0}^{\theta _{n}}d\theta _{n-1}\cdots \int_{0}^{\theta _{2}}d\theta_{1}
\prod_{\ell =1}^{n}\tilde{f}_{\lambda _{\ell }}(\theta _{\ell })
\label{atheta}
\end{equation}%
with $\tilde{f}_{\lambda _{\ell }}(\theta _{\ell })\equiv G(\theta_{\ell })f_{\lambda _{\ell }}(\theta _{\ell })$, where
\begin{eqnarray}
\tilde{f}_{0}&=&\sum_{k=0}^{\infty }\sum_{q=-1,1}g_{k,q}\sin[2k(N_{2}+1)\theta +q \theta ],  \label{f00} \\
\tilde{f}_{z}&=&\sum_{k=0}^{\infty }\sum_{q=-1,1}d_{k,q}^{z}\cos [(2k+1)(N_{2}+1)\theta+q\theta ].  \label{fzz}
\end{eqnarray}
When the inner decoupling order $N_{1}$ is even, 
\begin{eqnarray}
\tilde{f}_{x}&=&\sum_{k=0}^{\infty }\sum_{q=-1,1}d_{k,q}^{x}\sin [2k(N_{2}+1)\theta +q \theta ],\label{fxe}\\
\tilde{f}_{y}&=&\sum_{k=0}^{\infty }\sum_{q=-1,1}d_{k,q}^{y}\cos [(2k+1)(N_{2}+1)\theta +q\theta ], \label{fye} 
\end{eqnarray}
while if it is odd, 
\begin{eqnarray}
\tilde{f}_{x} &=&\sum_{k=0}^{\infty }\sum_{q=-1,1}d_{k,q}^{x}\cos [2k(N_{2}+1)\theta +q \theta ],  \label{fxo}\\
\tilde{f}_{y}&=&\sum_{k=0}^{\infty }\sum_{q=-1,1}d_{k,q}^{y}\sin [(2k+1)(N_{2}+1)\theta +q\theta ].\label{fyo}
\end{eqnarray}
Observe that all the integrands of $a_{\lambda _{n}\cdots \lambda _{1}}$ are composed of sums of either purely cosine functions or purely sine functions, i.e., none of the integrands is a mixed sum.  This fact is key to our ability to perform the nested integral, as we show next.


\subsection{Procedure to evaluate nested multiple integrals with integrands being either a cosine series or a sine series}
\label{sec: steps}

Suppose that, up to an order $N$ which depends on $N_1$ and $N_2$, all the normalized QDD$_{N_{1}, N_{2}}$ coefficients $a_{\lambda_{n}\cdots \lambda _{1}}$ [multiple nested integral Eq.~\eqref{a}] can be  reduced to a single integral as either
\begin{eqnarray}
&&\sum_{P\in \mathbb{Z}} \int_{0}^{\pi}\sin[\,P\,\theta\,] d\theta  \qquad \text{    or} \label{sin} \\
&&\sum_{P\in \mathbb{Z}\smallsetminus 0} \int_{0}^{\pi}\cos [\,P\,\theta\,] d\theta \label{cos}
\end{eqnarray}
where we omit prefactors for simplicity. We shall show in the following subsections that this form arises in the evaluation of the QDD$_{N_1 N_2}$ coefficients.

Moreover, we shall show in the following subsections that all $a_{\lambda_{n}\cdots \lambda _{1}=Z}$ coefficients with order  $n\leq N$ are of the form of Eq.~\eqref{cos}, and hence that $a_{\lambda_{n}\cdots \lambda _{1}=Z}$ vanishes since 
\begin{equation}
\sum_{P\in \mathbb{Z}\smallsetminus 0} \sin [\,P\,\theta\,]|^{\pi}_{0}=0
\label{sin pi}
\end{equation} 
after performing the last integral.  Therefore, the dephasing errors $\sigma_{Z}$ can be eliminated at least up to a remaining error of $\mathcal{O}(T^{N+1})$.

We first note that regardless of the integration limits, all $a_{\lambda_{n}\cdots \lambda _{1}}$ coefficients can be viewed as one integral nested with one order lower ($n-1^{th}$ order) coefficient, 
\begin{equation}
a_{\lambda_{n}\cdots \lambda _{1}}= \int_0^\pi d\theta_n \tilde{f}_{\lambda_{n}}(\theta_n)\,  a^{\theta_{n}}_{\lambda_{n-1}\cdots \lambda _{1}}
\label{nest}
\end{equation}
where the superscript $\theta_{n}$ indicates that the upper integration limit of $a^{\theta_{n}}_{\lambda_{n-1}\cdots \lambda _{1}}$ is $\theta_{n}$, not $\pi$. Now assume that all the $n-1^{th}$ order coefficients $a_{\lambda_{n-1}\cdots \lambda _{1}}$ are of the form of Eq.~\eqref{sin} or Eq.~\eqref{cos}. Then one could just proceed to the next order by substituting Eq.~\eqref{sin} or Eq.~\eqref{cos} (with upper integration limits $\pi$ replaced by $\theta_{n}$ for the $n-1^{th}$ order coefficient $a^{\theta_{n}}_{\lambda_{n-1}\cdots \lambda _{1}}$) into Eq.~\eqref{nest}, with the $n$-level nested integral having been reduced to a two-fold nested integral. Therefore, under the assumption above, two-fold nested integrals are the basic units for evaluating multiple nested integrals.

It follows from the result in the previous subsection that all $\tilde{f}_{\lambda_{n}}$ are sums of purely sine or purely cosine functions. Combining this with Eq.~\eqref{nest} and the assumption that all the $n-1^{th}$ order coefficients $a_{\lambda_{n-1}\cdots \lambda _{1}}$ are of the form of Eq.~\eqref{sin} or Eq.~\eqref{cos}, there are only four possible types of two-fold nested integrals, which are presented on the left hand sides of Eq.~\eqref{twointegrals}. The results, on the right, follow simply from evaluation of the $\theta_{n-1}$ integrals, followed by application of the product-to-sum trigonometric formulas~\eqref{sinsin}-\eqref{coscos}.
\bes
\label{twointegrals}
\begin{eqnarray}
&&\int  d\theta_{n} \sin[p_{s}\theta_{n}]\int_{0}^{\theta_{n}}d\theta_{n-1}\cos[P_{c}\theta_{n-1}] \sim \notag \\
&&\qquad \int d\theta_{n} \cos [(p_{s}\pm P_{c})\theta_{n} ]\\
&&\int  d\theta_{n}\cos[p_{c}\theta_{n}]\int_{0}^{\theta_{n}}d\theta_{n-1}\sin[P_{s}\theta_{n-1}]\sim \notag \\
&&\qquad \int d\theta_{n}\cos [(p_{c}\pm P_{s})\theta_{n} ]\\
&& \int  d\theta_{n}\cos [p_{c}\theta_{n}] \int_{0}^{\theta_{n}} d\theta_{n-1} \cos[P_{c}\theta_{n-1}] \sim \notag \\
&&\qquad \int d\theta_{n} \sin [( p_{c}\pm P_{c})\theta_{n}]\\ 
&&\int  d\theta_{n}\sin[p_{s}\theta_{n}]\int_{0}^{\theta_{n}}d\theta_{n-1}\sin[P_{s}\theta_{n-1}]\sim \notag \\
&&\qquad \int d\theta_{n}\sin [(p_{s}\pm P_{s})\theta_{n} ] 
\end{eqnarray}
\ees
where the $\pm$ symbol is shorthand for, e.g., $\int d\theta_{n} \cos [(p_{s}\pm P_{c})\theta_{n} ] \equiv \int d\theta_{n} \cos [(p_{s}+ P_{c})\theta_{n} ] + \int d\theta_{n} \cos [(p_{s}- P_{c})\theta_{n} ]$, and where we have omitted irrelevant prefactors in front of all integrals.

Note that the $\cos$ integrands on the right hand side of Eq.~\eqref{twointegrals}. will yield $1$ if their arguments happen to vanish. This conflicts with the requirement of Eq.~\eqref{cos}, and would prevent us from proving that $a_{\lambda_{n}\cdots \lambda _{1}=Z}$ vanishes. Likewise, in order to proceed to the next order, say order $n$,  none of the $n-1^{\rm th}$ order coefficients $a^{\theta_{n}}_{\lambda_{n-1}\cdots \lambda _{1}}$ in Eq.~\eqref{nest} may contain constant terms when expressed as a single integral of a cosine series.
The reason that a constant is problematic is that it behaves differently from a cosine function under integration. The integral of a cosine function with non-zero argument gives rise to a sine function, but the integral of a constant gives rise to a linear function. Therefore, if the integrand is a cosine series including a constant term,  then after integration the result will not be a pure sine series any more. Furthermore, the problem cannot be resolved by carrying out the next integral. On the other hand, one need not worry about sine functions because sine functions with arbitrary angles will always result in cosine functions after integration.  

Therefore, proceeding from $n-1^{\rm th}$ order to $n^{\rm th}$ order, suppose none of  $n-1^{\rm th}$ order coefficients $a^{\theta_{n}}_{\lambda_{n-1}\cdots \lambda _{1}}$  contain a constant term. From Eq.~\eqref{twointegrals}, due to the product-to-sum trigonometric formula, the problematic constant term will be generated when the new resulting argument $p_{s}\pm P_{c}$ in the cosine functions happens to vanish. When this happens to any one of the $n^{th}$ order coefficients $a_{\lambda_{n}\cdots \lambda _{1}}$, there is no advantage, when using our proof method, in proceeding to the $n+1^{th}$ order; this order is where the \red{cosine arguments} may start to be zero, and hence it sets a lower bound on the suppression order of the pure dephasing term.


\subsection{The suppression ability of the outer $X$-type UDD$_{N_{2}}$ sequence when $N_{1}$ is even} 

Let us define four function types we shall encounter in our proof.

\begin{mydef} 
$c^{n}_{\rm odd}$, $c^{n}_{\rm even}$, $\zeta^{n}_{\rm even}$, and $\zeta^{n}_{\rm odd}$ function types. Let $k,q \in \mathbb{Z}$ with $k$ arbitrary and $|q|\leq n$.\\
A $c^{n}_{\rm odd}$-type function is an arbitrary linear combination of $\cos[(2k+1)(N_{2}+1)\theta + q \theta ]$  terms.\\
A $c^{n}_{\rm even}$-type function is an arbitrary linear combination of $\cos[2k(N_{2}+1)\theta + q \theta ]$ terms.\\
A $\zeta^{n}_{\rm even}$-type function is an arbitrary linear combination of $\sin[2k(N_{2}+1)\theta + q \theta]$ terms.\\
A $\zeta^{n}_{\rm odd}$-type function is an arbitrary linear combination of $\sin[(2k+1)(N_{2}+1)\theta + q \theta]$ terms.
\label{def}
\end{mydef}

When $n\leq N_{2}$ we have $(2k+1)(N_{2}+1)+ q \neq 0$. Therefore, by definition,  all $c^{n}_{\rm odd}$-type functions will in this case have no constant $1$ (the problematic term). The $c^{n}_{\rm even}$-type functions are allowed to have a constant term.

From Eqs.~\eqref{f00}-\eqref{fye}, for even inner decoupling order $N_{1}$, there are only two kinds of integrands: $\widetilde{f_{0}}$ and $\widetilde{f_{x}}$ are $\zeta^{1}_{\rm even}$-type functions while  $\widetilde{f_{z}}$ and  $\widetilde{f_{y}}$ are $c^{1}_{\rm odd}$-type functions which, as we just remarked, do not have the constant $1$ term. Therefore, it immediately follows from Eqs.~\eqref{cos} and \eqref{sin pi} that the first order normalized QDD$_{N_{1}, N_{2}}$ coefficients $a_{Z}=\int_{0}^{\pi}\widetilde{f_{z}}\, d\theta$ and $a_{Y}=\int_{0}^{\pi}\widetilde{f_{y}}\, d\theta$ vanish.

Next, let us consider the second order terms (two-fold nested integrals), as in Eq.~\eqref{twointegrals}. We introduce a binary operation $\odot$ which (1) evaluates the first integrand, (2) multiplies the outcome with the second integrand, (3) applies the appropriate product-to-sum trigonometric formula. Substituting the $c^{1}_{\rm odd}$ or $\zeta^{1}_{\rm even}$-type functions into Eq.~\eqref{twointegrals}, we then have 
\bes
\label{twoint N1 even}
\begin{eqnarray}
&&\zeta^{1}_{\rm even} \odot c^{1}_{\rm odd} = c^{2}_{\rm odd} \\
&&c^{1}_{\rm odd} \odot \zeta^{1}_{\rm even}  = c^{2}_{\rm odd} \\
&&c^{1}_{\rm odd} \odot c^{1}_{\rm odd} = \zeta^{2}_{\rm even} \\
&&\zeta^{1}_{\rm even} \odot \zeta^{1}_{\rm even}  = \zeta^{2}_{\rm even}
\end{eqnarray}
\ees
where we omitted the second integration symbol.

If we disregard the $n$ superscript  of $c^{n}_{\rm odd}$ and $\zeta^{n}_{\rm even}$ in Eq.~\eqref{twoint N1 even}, the set $\{\zeta_{\rm even},c_{\rm odd}\}$ constitutes the abelian group $Z_{2}$ under the binary operation $\odot$, with the identity element $\zeta_{\rm even}$.

On the other hand, the superscript of the resulting function, $c^{2}_{\rm odd}$ or $\zeta^{2}_{\rm even}$ in Eq.~\eqref{twoint N1 even}, is just the sum of the superscripts of the first and second integrands ($c^{1}_{\rm odd}$ or $\zeta^{1}_{\rm even}$). Accordingly,  the binary operation $\odot$ acts as integer addition for the superscript $n$.

Let us now consider the $n$-fold nested integral implied by Eq.~\eqref{nest}. Because of the closure property of the group $Z_{2}$, integer addition of the superscripts $n$ of $c^{n}_{\rm odd}$, and $\zeta^{n}_{\rm even}$, and the fact that no $c^{n}_{\rm odd}$-type function with  $n\leq N_{2}$ contains the constant $1$, we can conclude that such an $n$-fold nested integral  with  $n\leq N_{2}$ and with each integrand being either $c^{1}_{\rm odd}$-type or $\zeta^{1}_{\rm even}$-type functions  can be reduced to be either $\int c^{n}_{\rm odd} d\theta_{n}$ or $\int \zeta^{n}_{\rm odd} d\theta_{n}$.
 
More specifically, note that in Eq.~\eqref{twoint N1 even} the first two lines have an odd number of $c^{1}_{\rm odd}$ functions and result in $c^{2}_{\rm odd}$, while the last two lines have an even number of $c^{1}_{\rm odd}$ functions and result in $\zeta^{2}_{\rm even}$. When we continue the nesting process using these rules, the odd or even property is maintained while the $n$ superscript grows by one unit each time. In other words, due to $Z_{2}$ group multiplication rules [Eq.~\eqref{twoint N1 even} without the superscripts $n$], we have the following lemma:
\begin{mylemma}
Provided $n \leq N_{2}$, all $n$-fold nested integrals, with each integrand being  either a $c^{1}_{\rm odd}$-type or $\zeta^{1}_{\rm even}$-type function, can be written as 
\begin{enumerate}
 \item  $\int\, c^{n}_{\rm odd}\, d\theta_{n}$ if there is an  {\em odd} total number of $c^{1}_{\rm odd}$-type integrands in the $n$-fold nested integral,
 \item $\int \, \zeta^{n}_{\rm even} \, d\theta_{n}$ if there is an {\em even} total number of $c^{1}_{\rm odd}$-type integrands in the $n$-fold nested integral. 
\end{enumerate} 
\label{lem:N1 even}
\end{mylemma}

Next, let us determine the parity of the number of $c^{1}_{\rm odd}$-type functions appearing in the QDD$_{N_1 N_2}$ coefficients. Consider, e.g., $a_{\lambda_{n}\cdots \lambda _{1}=Z}$. Recall that $\widetilde{f_{0}}$ and $\widetilde{f_{x}}$ are $\zeta^{1}_{\rm even}$-type functions while  $\widetilde{f_{z}}$ and  $\widetilde{f_{y}}$ are $c^{1}_{\rm odd}$-type functions. Consulting the last column of Part 1 of Table \ref{table:pauli} (the second $a_{\lambda_{n}\cdots \lambda _{1}=Z}$ column), we see that there is an odd number of $f_x$ ($\zeta^{1}_{\rm even}$) and $f_y$ ($c^{1}_{\rm odd}$), and an even number of $f_z$ ($c^{1}_{\rm odd}$). Therefore, in this case, we have an odd$+$even$=$odd number of $c^{1}_{\rm odd}$-type functions in $a_{\lambda_{n}\cdots \lambda _{1}=Z}$. Similarly, consulting all other columns of Part 1 of Table \ref{table:pauli},  it turns out that all possible combinations generating $a_{\lambda_{n}\cdots \lambda _{1}=Z}$ or $a_{\lambda_{n}\cdots \lambda _{1}=Y}$  contain an odd number of $c^{1}_{\rm odd}$-type functions. It now follows from Lemma \ref{lem:N1 even} and then Eq.~\eqref{sin pi} that all $a_{\lambda_{n}\cdots \lambda _{1}=Z}=a_{\lambda_{n}\cdots \lambda _{1}=Y}=0$ if the order $n\leq N_{2}$. [Note that this counting argument is unaffected by the move from $f$ to $\tilde{f}$, since this move was due to a change of integration variables---see Eq.~\eqref{atheta}.]

In conclusion, the inner UDD$_{N_{1}}$ sequences with even order $N_{1}$ do not affect the suppression effect of the outer UDD$_{N_{2}}$ sequence, i.e., the $\sigma_Z$  error is always removed up to the expected order $N_{2}$ when the order $N_{1}$ of the inner sequence is even. This proves the first row of the $\sigma_{Z}$ part of Table \ref{table:qdd}.

In addition, we have just shown that  when the inner order $N_{1}$ is even,  the outer $X$-type UDD$_{N_{2}}$ sequence also eliminates the $\sigma_{Y}$  error up to the outer decoupling order $N_{2}$. Since we have shown in Sec. \ref{sec: ax and ay} that the $\sigma_{Y}$ error is suppressed to order $N_{1}$ when $N_{2}$ is even, or $N_{1}+1$  when $N_{2}$ is odd, one can conclude that when the inner order $N_{1}$ is even, $\sigma_{Y}$ is suppressed to order $\max[N_{1},N_{2}]$ when $N_{2}$ is even, and to order $\max[N_{1}+1,N_{2}]$ when $N_{2}$ is odd. This completes the proof of the $\sigma_{Y}$ part in Table \ref{table:qdd}.


\subsection{The suppression ability of the outer $X$-type UDD$_{N_{2}}$ sequence when $N_{1}$ is odd}

The main difference between the analysis in this subsection and the previous one is that $\tilde{f}_x$ and $\tilde{f}_y$ are interchanged in terms of which function is cosine or sine---see Eqs.~\eqref{fxe}-\eqref{fyo}. 

Also, note that, from Eqs.~\eqref{f00}, \eqref{fzz}, \eqref{fxo}, and \eqref{fyo}, for odd inner decoupling order $N_{1}$, $\tilde{f_{0}}$ is a $\zeta^{1}_{\rm even}$-type function, $\tilde{f_{z}}$ is a $c^{1}_{\rm odd}$-type function, $\tilde{f_{x}}$ is a $c^{1}_{\rm even}$-type, and $\tilde{f_{y}}$ is a $\zeta^{1}_{\rm odd}$-type. We shall use these facts throughout this subsection.

Our procedure is to start from the first order QDD coefficients, then the second order, and finally the general, $n$th order.

\subsubsection{The first order terms $a_{\lambda _{1}}$}

It immediately follows from the fact that $\tilde{f_{z}}$ is a $c^{1}_{\rm odd}$-type function and from Eq.~\eqref{cos} that $a_{Z}=\int_{0}^{\pi}\tilde{f_{z}}\, d\theta=0$.  It also immediately follows from the function types that $\tilde{f_{0}}$ and $\tilde{f_{y}}$ are of the form of Eq.~\eqref{sin}. The only function that deserves special attention is $\tilde{f_{x}}$. 

As discussed in Sec. \ref{sec: steps}, in order to proceed to second order, none of the modulation functions can contain a constant $1$ term. However, Definition \ref{def} allows $c^{1}_{\rm even}$-type functions to have such a term. Accordingly, before applying Eq.~\eqref{nest} to the second order case, we should check whether $\tilde{f_{x}}(\theta )$ has a constant term. 

Suppose $\tilde{f_{x}}(\theta )$ has a constant $1$ term and then separate the constant $1$ from the other cosine functions with non-zero arguments as follows,
\begin{equation}
\tilde{f_{x}}(\theta )=\sum_{p \neq 0} d_{p}\cos[\,p\,\theta\,] + r
\end{equation}
where $p=2k(N_{2}+1)\pm q$ with $|q|\leq 1$  an integer, $r$  a coefficient of the constant $1$ term, and $d_{p}$ coefficients of $\cos[\,p\,\theta\,]$. Then the first order normalized QDD$_{N_1 N_2}$ coefficient of the $\sigma_{X}$ error reads
\begin{eqnarray}
a_{X}&=&\int_{0}^{\pi}\tilde{f_{x}}(\theta )\,d\theta \notag\\
&=& \sum_{p \neq 0}d_{p}\sin[\,p\,\theta\,]|_{\theta=0}^{\theta=\pi}+r\theta|_{\theta=0}^{\theta=\pi} \notag\\
&=& 0 + r\pi
\label{ax}
\end{eqnarray}
Now, since we already proved in Section \ref{sec: ax and ay} that $a_{\lambda_{n}\dots\lambda_{1}=X}=0$ for $n\leq N_{1}$, and since in the first order case $n=1$ and hence $n\leq N_{1}$ always holds, it follows that $r=0$. 
Therefore, $\tilde{f}_{x}$ does not contain a constant $1$ term.

\begin{table*}[htbp]
\begin{tabular}{|c||c|c|c|c|}\hline
$\odot$             & $a_{I}:\,\zeta^{1}_{\rm even}$ & $a_{X}:\,c^{1}_{\rm even}$     & $a_{Y}:\,\zeta^{1}_{\rm odd}$ & $a_{Z}:\,c^{1}_{\rm odd}$\\ \hline \hline
$a_{I}:\,\zeta^{1}_{\rm even}$ & $a_{II=I}:\, \zeta^{2}_{\rm even}$ & $a_{IX=X}:\,c^{2}_{\rm even}$     & $a_{IY=Y}:\,\zeta^{2}_{\rm odd}$ & $a_{IZ=Z}:\,c^{2}_{\rm odd}$\\ \hline
$a_{X}:\,c^{1}_{\rm even}$     & $a_{XI=X}:\,c^{2}_{\rm even}$     & $a_{XX=I}:\,\zeta^{2}_{\rm even}$ & $a_{XY=Z}:\,c^{2}_{\rm odd}$     & $a_{XZ=Y}:\,\zeta^{2}_{\rm odd}$\\ \hline
$a_{Y}:\,\zeta^{1}_{\rm odd}$  & $a_{YI=Y}:\,\zeta^{2}_{\rm odd}$  & $a_{YX=Z}:\,c^{2}_{\rm odd}$      & $a_{YY=I}:\,\zeta^{2}_{\rm even}$& $a_{YZ=X}:\,c^{2}_{\rm even}$\\ \hline
$a_{Z}:\,c^{1}_{\rm odd}$      & $a_{ZI=Z}:\,c^{2}_{\rm odd}$      & $a_{ZX=Y}:\,\zeta^{2}_{\rm odd}$  & $a_{ZY=X}:\,c^{2}_{\rm even}$    & $a_{ZZ=I}:\,\zeta^{2}_{\rm even}$ \\ \hline
\end{tabular}
\caption{The group structure associated with the second order QDD$_{N_1 N_2}$ coefficient $a_{\lambda_{2}\lambda _{1}}=a_{\lambda_{2}}\odot a^{\theta_{2}}_{\lambda_{1}}$.}
\label{table:a2 N1 odd}\centering
\end{table*}

In summary, now that we have shown that $a_{\lambda_1=Z}=0$ and that all the first order normalized QDD$_{N_{1}, N_{2}}$ coefficients $a_{\lambda _{1}}$ are of the form of either Eq.~\eqref{sin} or Eq.~\eqref{cos}, 
we can proceed to the second order case.


\subsubsection{The second order terms $a_{\lambda_{2}\lambda _{1}}$ }

From Eq.~\eqref{nest}, the second order normalized QDD$_{N_{1}, N_{2}}$ coefficients  is
\begin{equation}
a_{\lambda_{2}\lambda _{1}}= \int_0^\pi d\theta_2 \tilde{f}_{\lambda_{2}}(\theta_2)\,  a^{\theta_{2}}_{\lambda_{1}}.
\end{equation}
After additionally applying the trigonometric product-to-sum transformation, and using the operation $\odot$ defined above Eq.~\eqref{twoint N1 even}, we can write
\beq
a_{\lambda_{2}\lambda _{1}}=a_{\lambda_{2}}\odot a^{\theta_{2}}_{\lambda_{1}}
\label{a12}
\eeq
Next, in Eq.~\eqref{a12}, let us substitute $\zeta^{1}_{\rm even}$ into the integrand of $a_{I}$, $c^{1}_{\rm even}$ into $a_{X}$, $\zeta^{1}_{\rm odd}$ into $a_{Y}$, and $c^{1}_{\rm odd}$ into $a_{Z}$. The resulting set of all $a_{\lambda_{2}\lambda _{1}}$ can be arranged into a multiplication table, Table~\ref{table:a2 N1 odd},
where the entries in the top row are the types of the first integrand and the entries in the left-most column are the types of the second integrand. The remaining entries are the results of applying the binary operation $\odot$ between the elements of the first row and column.

From Table ~\ref{table:a2 N1 odd},  the superscript of the resulting function is again the sum of the superscripts of the first and second integrands. Hence the binary operation $\odot$ again acts as integer addition for the superscripts. 
Moreover, disregarding the superscripts $n$, Table ~\ref{table:a2 N1 odd} shows that the set $\{\zeta_{\rm even},\, c_{\rm even},\,  \zeta_{\rm odd},\,  c_{\rm odd}\}$ forms the abelian Klein four-group, i.e., the $Z_{2}\times Z_{2}$ group, under the binary operation $\odot$.  The key observation from Table ~\ref{table:a2 N1 odd} is that the algebra of the subscripts $\lambda_{2}\lambda _{1}$ of $a_{\lambda_{2}\lambda _{1}}$ works as the \textit{Pauli algebra  without the anti-commutativity property}, which is isomorphic to the Klein four-group algebra by mapping the identity $I$ to $\zeta_{\rm even}$, $X$ to $c_{\rm even}$, $Y$ to $\zeta_{\rm odd}$, and  $Z$ to $c_{\rm odd}$.

Accordingly, the results of Table ~\ref{table:a2 N1 odd} can be summarized as follows,
\bes
\label{zn2}
\begin{eqnarray}
a_{\lambda_{2}\lambda _{1}=Z}&=&\int_{0}^{\pi}\, c^{2}_{\rm odd} \, d\theta  \\
a_{\lambda_{2}\lambda _{1}=X}&=&\int_{0}^{\pi}\, c^{2}_{\rm even} \, d\theta \\
a_{\lambda_{2}\lambda _{1}=Y}&=&\int_{0}^{\pi}\, \zeta^{2}_{\rm odd} \, d\theta \\
a_{\lambda_{2}\lambda _{1}=I}&=&\int_{0}^{\pi}\, \zeta^{2}_{\rm even} \, d\theta. \label{0n2} 
\end{eqnarray}
\ees
We can conclude that $a_{\lambda_{2}\lambda _{1}=Z}=0$ if $N_{2} \geq 2$, since then (by definition) $c^{2}_{\rm odd}$ does not contain a constant $1$ term. 

We have already proved in Section \ref{sec: ax and ay} that $a_{\lambda_{2}\lambda _{1}=X}=0$ if $N_{1}\geq 2$. By the same argument as Eq.~\eqref{ax}, this implies that the integrand $c^{2}_{\rm even}$ does not have a constant $1$ term if $N_{1}\geq 2$. 

In order to proceed to the next order none of the integrands may contain a constant. 
Therefore, our results show that if $N_{1},N_2 \geq 2$, one can indeed proceed to the next order. On the other hand, if $N_1=N_2=1$ we can only conclude that $a_{\lambda _{1}=Z}=0$, while
if $N_1=1$ and $N_2 \geq 2$, we can only conclude that $a_{\lambda _{1}=Z}=a_{\lambda_{2}\lambda _{1}=Z}=0$, but not that the third or higher order $Z$-type QDD coefficients are zero.


\subsubsection{The $n^{th}$ order terms $a_{\lambda_{n}\cdots \lambda _{1}}$}

The procedure we described for the first and second orders applies to higher orders, until one reaches the order $N$ where some resulting integrands begin to include constant $1$ terms. 

To obtain the $n$th order QDD$_{N_1 N_2}$ coefficients we proceed by induction on $n$. We have already established the case of $n=1$ and $n=2$. Let us assume that 
\bes
\label{azn-1}
\begin{eqnarray}
a_{\lambda_{n}\cdots \lambda _{1}=Z}&=&\int_{0}^{\pi}\, c^{n}_{\rm odd} \, d\theta \label{azn}\\
a_{\lambda_{n}\cdots \lambda _{1}=X}&=&\int_{0}^{\pi}\, c^{n}_{\rm even} \, d\theta \label{axn}\\
a_{\lambda_{n}\cdots \lambda _{1}=Y}&=&\int_{0}^{\pi}\, \zeta^{n}_{\rm odd} \, d\theta \label{ayn}\\
a_{\lambda_{n}\cdots \lambda _{1}=I}&=&\int_{0}^{\pi}\, \zeta^{n}_{\rm even} \, d\theta. \label{a0n}
\ignore{
a_{\lambda_{n-1}\cdots \lambda _{1}=Z}&=&\int_{0}^{\pi}\, c^{n-1}_{\rm odd} \, d\theta \\
a_{\lambda_{n-1}\cdots \lambda _{1}=X}&=&\int_{0}^{\pi}\, c^{n-1}_{\rm even} \, d\theta \label{axn-1}\\
a_{\lambda_{n-1}\cdots \lambda _{1}=Y}&=&\int_{0}^{\pi}\, \zeta^{n-1}_{\rm odd} \, d\theta \label{ayn-1}\\
a_{\lambda_{n-1}\cdots \lambda _{1}=I}&=&\int_{0}^{\pi}\, \zeta^{n-1}_{\rm even} \, d\theta , \label{a0n-1}
}
\end{eqnarray}
\ees
where none of these integrals contains a constant $1$ term in their integrand,
and prove that the same integrand form holds for $n+1$ (but not necessarily that there is no constant $1$). Indeed, using the definition of the $\odot$ operation and Eq.~\eqref{nest}, we have
\beq
a_{\lambda_{n+1}\cdots \lambda _{1}} = a_{\lambda_{n+1}}\odot a_{\lambda_{n}\cdots \lambda _{1}}^{\theta_{n+1}}
\eeq
Due to the induction assumption [Eq.~\eqref{azn-1}] the situation is now identical to the one we analyzed for $n=2$, in particular in Eq.~\eqref{zn2}. Therefore Eq.~\eqref{azn-1} holds with $n$ replaced by $n+1$.
\ignore{
we have, for any $n^{th}$ order $a_{\lambda_{n}\cdots \lambda _{1}}$ with $n \leq N$,
\begin{eqnarray}
a_{\lambda_{n}\cdots \lambda _{1}=Z}&=&\int_{0}^{\pi}\, c^{n}_{\rm odd} \, d\theta \label{azn}\\
a_{\lambda_{n}\cdots \lambda _{1}=X}&=&\int_{0}^{\pi}\, c^{n}_{\rm even} \, d\theta \label{axn}\\
a_{\lambda_{n}\cdots \lambda _{1}=Y}&=&\int_{0}^{\pi}\, \zeta^{n}_{\rm odd} \, d\theta \label{ayn}\\
a_{\lambda_{n}\cdots \lambda _{1}=I}&=&\int_{0}^{\pi}\, \zeta^{n}_{\rm even} \, d\theta. \label{a0n}
\end{eqnarray}
}

This can also be understood without induction as being due to the isomorphism between the set $\{\zeta_{\rm even},\, c_{\rm even},\,  \zeta_{\rm odd},\,  c_{\rm odd}\}$ and the set $\{I,\,X,\,Y,\,Z\}$ (the subscripts of $a_{\lambda_{n}\cdots \lambda _{1}}$), and the addition of superscripts under the $\odot$ operation.

To figure out up to which order $N$ Eq.~\eqref{azn-1} holds, one must examine when the $ c^{n}_{\rm odd}$ or  $c^{n}_{\rm even}$-type functions begin to have constant $1$ terms. The $c^{n}_{\rm odd}$-type functions will by definition not contain constant $1$ terms until order $n=N_{2}+1$. On the other hand, due to Eq.~\eqref{axn} and $a_{\lambda_{n}\cdots \lambda _{1}=X}=0$ for $n \leq N_{1}$ (proven in Sec. \ref{sec: ax and ay}),  $c^{n}_{\rm even}$ in Eq.~\eqref{axn} is guaranteed to have no constant $1$ term until order $n=N_{1}+1$, by a similar argument as that leading to Eq.~\eqref{ax}. In conclusion,

\begin{mylemma}
For QDD$_{N_{1} N_{2}}$ with odd $N_{1}$, all $n$th order normalized QDD$_{N_{1} N_{2}}$ coefficients  $a_{\lambda_{n}\cdots \lambda _{1}}$ with $n \leq \min[N_{1}+1,N_{2}+1]$ can be written as  Eq.~\eqref{azn-1}, and none of the integrands in Eq.~\eqref{azn-1} contain a constant $1$ term when $n \leq \min[N_{1},N_{2}]$. 
\label{lem:N1 odd}
\end{mylemma}

It follows immediately from Lemma~\ref{lem:N1 odd} and Eq.~\eqref{sin pi} that the first  $\min[N_{1},N_{2}]$ orders of $a_{\lambda_{n}\cdots \lambda _{1}=Z}$ vanish. 
However, in fact we can show more, namely that $a_{\lambda_{n}\cdots \lambda _{1}=Z}=0$ for all $n \leq \min[N_{1}+1,N_{2}]$. Suppose that $N_1 < N_2$ and consider the special case $n=N_1+1$. In this case it follows from Lemma~\ref{lem:N1 odd} that $a_{\lambda_{N_1+1}\cdots \lambda _{1}=Z} = \int_{0}^{\pi}\, c^{N_1+1}_{\rm odd} \, d\theta$; the argument of the function $c^{N_1+1}_{\rm odd}$ is $(2k+1)(N_2+1)\theta+q\theta$, and $|q| \leq N_1+1$. Since $N_1<N_2$ this argument cannot vanish, and it follows that $a_{\lambda_{N_1+1}\cdots \lambda _{1}=Z}=0$. In conclusion, $a_{\lambda_{n}\cdots \lambda _{1}=Z}=0$ for all $n \leq \min[N_{1}+1,N_{2}]$, which proves the last row in Table \ref{table:qdd}.

In summary, if the inner decoupling order $N_{1}$ is odd and $N_{2}\leq N_{1}+1$, the outer UDD$_{N_{2}}$ sequence always suppresses the dephasing error $Z$ to the expected decoupling order $N_{2}$, as then $\min[N_{1}+1,N_{2}]=N_{2}$. In contrast, if the inner decoupling order $N_{1}$ is odd and $N_{2}> N_{1}+1$, the outer UDD$_{N_{2}}$ sequence  suppresses the dephasing error $Z$ (at least) up to order $N_{1}+1$, which may be smaller than the expected outer decoupling order $N_{2}$. Thus, if the order of inner level UDD$_{N_{1}}$ sequence is odd, this \red{may} inhibit the suppression ability of the outer  UDD$_{N_{2}}$ sequence.


\section{comparison between our theoretical bounds and numerical results}
\label{sec: numerical and theory}

In Ref.~\cite{QuirozLidar:11} the QDD sequence was analyzed numerically and the scaling of the single-axis errors was determined on the basis of simulations, for $N_1$ and $N_2$ in the range $\{1,\dots,24\}$. These simulations are in complete agreement with our analytically bounds for $n_{x}$ and $n_{y}$, as given in Table~\ref{table:qdd}.  They are also in complete agreement with our bound for $n_z$ when $N_1$ is even. Thus we can conclude that it is likely that our bounds are in fact tight in these cases. There is, however, one discrepancy: when $N_1$ is odd our analytical bound yields $n_z = \min[N_1+1,N_2]$, while the numerical result found in Ref.~\cite{QuirozLidar:11} is $n_z = \min[2N_1+1,N_2]$. Thus, in this case our bound is not tight. We attribute this to the fact that the method we used in Sec. \ref{sec:outer} does not use the full information contained in the integrands, i.e., we discard all Fourier coefficients. Specifically, if $a_{\lambda_{n}\dots \lambda _{1}=Z}$ contains a constant term, namely, $\cos [P \theta ]$ with $P=0$, or does not end up in the form of Eq.~\eqref{cos},  it is still possible that $a_{\lambda_{n}\dots \lambda _{1}=Z}$ vanishes because a sum of non-zero terms could be zero when combined with the right Fourier coefficients. Thus, our method of analysis merely yields a lower bound on the decoupling order of the pure dephasing error. It is an interesting open problem to try to improve this bound so that it matches the numerical results of Ref.~\cite{QuirozLidar:11}.


\section{Summary and Conclusions}
\label{sec:conclusion}
 
The QDD sequence, introduced in Ref.~\cite{WestFongLidar:10}, is, to date, the most efficient pulse sequence known for suppression of single-qubit decoherence. In this work we provided a complete proof of the validity of this sequence, i.e., we proved its universality (independence of details of the environment) and performance. Our work complements an earlier proof \cite{WangLiu:11}, which was restricted to even order inner UDD sequences. However, our results go beyond a validity proof of QDD. For, in this work we also elucidated the dependence of single-axis error suppression on the orders $N_1$ and $N_2$ of the inner $X$-type and outer $Z$-type UDD sequences comprising QDD$_{N_1 N_2}$, respectively. Our results are stated in Theorem 1. Let us briefly summarize our method and main findings. 

Our general proof idea was to analyze the conditions under which, for each error type $\sigma_\lambda$, the $n$th order QDD$_{N_1 N_2}$ coefficients [Eq.~\eqref{U}] vanish. We used two complementary methods. In the first method, we expressed the QDD coefficients $a_{\lambda _{n}\cdots \lambda _{1}}$ in terms of UDD coefficients by splitting each of $a_{\lambda _{n}\cdots \lambda _{1}}$'s nested integrals into a sum of sub-integrals over normalized outer intervals. We were then able to conclude that $a_{\lambda _{n}\cdots \lambda _{1}=X}$ and $a_{\lambda _{n}\cdots \lambda _{1}=Y}$ vanish when $n\leq N_{1}$ due to the vanishing of the UDD$_{N_1}$ contributions.
\ignore{
We did so by considering first the case of $a_{\lambda _{n}\cdots \lambda _{1}=X}$ and $a_{\lambda _{n}\cdots \lambda _{1}=Y}$. We identified when these coefficients are proportional to vanishing UDD$_{N_1}$ contributions, which allowed us to conclude when they themselves vanish.}
For the $\sigma_Y$ error, still as part of the first method, we showed that an additional order vanishes due to a parity cancellation effect involving the outer sequence. However, this additional cancellation cannot be attributed to the vanishing of a corresponding UDD coefficient.
In the second method we considered the case of $a_{\lambda _{n}\cdots \lambda _{1}=Z}$, for which we provided an analysis based on the evaluation of integrals of trigonometric functions. We showed that their properties under \red{nested integration} can  be mapped to the Abelian groups $Z_2$ (for even $N_1$) and $Z_{2}\times Z_{2}$ (for odd $N_1$). Using this we provided a proof by induction for the vanishing of $a_{\lambda _{n}\cdots \lambda _{1}=Z}$, and, when $N_1$ is even, also for $a_{\lambda _{n}\cdots \lambda _{1}=Y}$.

The overall summary of our results is that $a_{\lambda _{n}\cdots \lambda _{1}=\lambda}=0$  $\forall n\leq N$, where $N$ is the decoupling order given in the last column of Table~\ref{table:qdd}. We now provide a recap of these results, including a semi-intuitive explanation based on the idea of  interference between the modulation functions.

Starting from the simplest case, we showed explicitly that independently of the order of the outer $X$-type sequence, the inner $Z$-type UDD$_{N_{1}}$ sequence always achieves its expected error suppression order, i.e., the  $\sigma_{X}$ and $\sigma_{Y}$ errors are suppressed to the inner decoupling order $N_{1}$. Since $\sigma_{X}$ errors commute with the pulses of the outer sequence they are not suppressed any further. 

The story is more complicated for the $\sigma_{Y}$ and $\sigma_{Z}$ errors, as they are both suppressed by the outer sequence. 

For the $\sigma_{Y}$ error, the parities of the inner and outer sequence orders cause the decoupling order to vary between $N_1$, $N_1+1$, and $N_2$. 
Consider first the even $N_2$  case.
An intuitive explanation for the corresponding parity effects is the following. For even $N_1$, the modulation functions $f_y$ and $f_z$ are {\em in phase}, namely both have a $\sin[(2k+1)(N_2+1)\theta]$ dependence [recall Eqs.~\eqref{afz} and \eqref{afy}]. The outer $X$-type UDD$_{N_{2}}$ sequence, with its $f_z$ modulation function, is then fully effective at eliminating the $\sigma_{Y}$ error, with the result that $\sigma_{Y}$ is eliminated to the expected decoupling order $\max[N_1,N_{2}]$. However, when $N_1$ is odd, $f_y$ has a $\cos[(2k+1)(N_2+1)\theta]$ dependence, which is $90$ degrees out of phase with $f_z$. In this case $f_y$ and $f_z$ interfere destructively with one another, and the outer sequence does not help to further suppress $\sy$. The result is that $\sigma_{Y}$ is only eliminated to order $N_1$.

Now consider the case of odd $N_2$. This case gives rise to the anomalous $N_1+1$ suppression order. The reason is that when $N_2$ is odd, the modulation function $f_z$ is odd with respect to the midpoint of the total sequence duration, while $f_x$ and $f_y$ are both even. It is this oddness of the outer sequence modulation function ($f_z$) which helps to suppress the error $\sy$ to one more order, due to a cancellation of terms with equal magnitude but opposite sign [Eq.~\eqref{odd-cancel}]. This gives rise to a cancellation to order $\max[N_1+1,N_2]$ when $N_1$ is even and \red{the inner sequence} does not interfere with the outer sequence, or to 
order $N_1+1$ when $N_1$ is odd and \red{the inner sequence does interfere with the outer sequence.}

Thus, suppose we fix $N_2$ so that it is even (odd) and greater than $N_1$ ($N_1+1$). We should then see the suppression order of $\sigma_Y$ switch between $N_1$ ($N_1+1$) and $N_2$, as $N_1$ is increased from $1$ to $N_2$, a phenomenon which was indeed observed in the numerical simulations of Ref.~\cite{QuirozLidar:11}.

If the inner order $N_{1}$ is even,  the outer $X$-type UDD$_{N_{2}}$ sequence always suppresses $\sigma_{Z}$ to the expected decoupling order $N_{2}$. This has the same intuitive origin as the $\sy$ case. Namely, for even $N_1$, $f_y$ and $f_z$ are in phase, i.e., both have a $\sin[(2k+1)(N_2+1)\theta]$ dependence, and so are able to suppress $\sz$ to the expected order.
\ignore{In other words, increasing the number of inner $Z$-type pulses does not help the suppression of the $\sigma_{Z}$ error.  
However, somewhat counterintuitively, when $N_1$ is odd and too small, the suppression of the $\sigma_{Z}$  error depends on the parity of the inner $Z$-type UDD$_{N_{1}}$ sequence, even though they commute.}
However, when $N_1$ is odd, the dependence of $f_y$ is $\cos[(2k+1)(N2+1)\theta$, which is $90$ degrees out of phase with $f_z$. Therefore again $f_y$ and $f_z$ interference destructively, and the outer sequence does not suppress the error $\sz$ to the expected order.

In more detail, if the inner order $N_{1}$ is odd and $N_{2}>N_{1}+1$, our proof method shows that the outer $X$-type UDD$_{N_{2}}$ sequence suppresses the  $\sigma_{Z}$ error at least to order $N_{1}+1$, which is less than the expected outer decoupling order $N_{2}$. Hence, if this lower bound is saturated, one can see a saturation effect in the decoupling order of $\sigma_{Z}$, which starts at $N_{2}=N_{1}+2$ when  we fix odd $N_{1}$ and increase $N_{2}$. Thus, odd $N_1$ can hinder the suppression ability of the outer sequence. 

The numerical results of Ref.~\cite{QuirozLidar:11} confirm that odd ${N_{1}}$ can hinder the suppression ability of the outer $X$-type UDD$_{N_{2}}$ sequence. However, the actual saturation effect in the decoupling order of $\sigma_{Z}$ begins at $N_{2}= 2N_{1}+\red{2}$, higher than our lower bound of $N_{2}=N_{1}+\red{2}$. A new method \red{may be} needed to explain the remaining vanishing orders from $N_{1}+2$ to $2N_{1}+1$.

The inhibitory effect of odd inner decoupling order $N_1$ disappears when $N_1$ is large enough. Specifically, whenever $N_{1} \geq N_{2}-1$ the outer $X$-type UDD$_{N_{2}}$ sequence suppresses $\sigma_{Z}$ to the expected decoupling order $N_{2}$. This makes intuitive sense because when $N_1$ is large enough the outer $X$-type UDD$_{N_{2}}$ sequence ``views" the effective Hamiltonian resulting from the inner $Z$-type UDD$_{N_{1}}$ sequence---which has time dependence $\mathcal{O}(T^{N_{1}+1\geq N_{2}})$---as time-independent relative to its ``error cancellation power" $\mathcal{O}(T^{N_{2}})$.   

Despite this complicated interplay between $N_1$ and $N_2$,  our proof yields the simple result that the QDD$_{N_1 N_2}$ sequence suppresses all single-qubit errors to an order $\geq \min[N_1,N_2]$. This matches the numerical results in \cite{QuirozLidar:11}, so that our bounds appear to be optimal in this regard. We conclude that to attain the highest order decoupling from the QDD$_{N_{1} N_{2}}$ sequence with ideal, zero-width pulses, one should use either an even order inner UDD sequence, or ensure that $N_1 \geq N_2-1$ if $N_1$ is odd.

A natural generalization of the work presented here is to NUDD with different sequence orders \red{\cite{KL:inprep}}. We look forward to experimental tests of the properties of the QDD$_{N_1 N_2}$ pulse sequence predicted in this work.

{\it Note added:} After this work was completed and while it was being written up for publication we became aware of a different, elegant proof of the universality of NUDD and in particular QDD \cite{JiangImambekov:10}. Our approach differs not only in methodology but also in providing a complete analysis of the single-axis errors.

\begin{acknowledgments}
We are grateful to \red{Gregory Quiroz, Gerardo Paz,  Jacob West, Stefano Pasini, and G\"{o}tz Uhrig} for very helpful
discussions. DAL acknowledges support from the NSF under Grants No. CHM-1037992 and CHM-924318.
Sponsored by United States Department of Defense.
The views and conclusions contained in this document are those of the authors and should not be interpreted as representing the official policies, either expressly or implied, of the U.S. Government.
\end{acknowledgments}

\appendix

\section{The form of $a_{\lambda _{n}\cdots \lambda _{1}}$ after outer interval  decomposition}
\label{app:form a}

We shall derive Eq.~\eqref{form a} by splitting each integral of $a_{\lambda _{n}\cdots \lambda _{1}}$ [Eq.~\eqref{a}] into a sum of sub-integrals over the normalized outer intervals $s_{j}$ in Eq.~\eqref{sj}. Since $a_{\lambda _{n}\dots\lambda _{1}}$ comprises a series of time-ordered, nested integrals, our procedure for decomposing $a_{\lambda _{n}\dots\lambda _{1}}$ is to split its nested integrals one by one, from $\eta^{(n)}$ to $\eta^{(1)}$. 

We call the sub-integral over the $j$th outer interval ``sub-integral-$j$''. Suppose the integral of the integration variable $\eta^{(\ell)}$ follows the sub-integral-$j^{(\ell+1)}$ of the previous variable $\eta^{(\ell+1)}$. By splitting the integral of $\eta^{(\ell)}$ with respect to the normalized outer intervals  and using Eq.~\eqref{f in out}, we have 
\bes
\begin{eqnarray}
&& \int_{0}^{\eta^{(\ell+1)}}f_{\lambda _{\ell}}(\eta^{(\ell)})\,d\eta^{(\ell)}\notag\\
&&=\sum_{j^{(\ell)}=1}^{j^{(\ell+1)}-1}f_{\beta_{\ell}}(j^{(\ell)})\int_{\eta_{j^{(\ell)}-1}}^{\eta_{j^{(\ell)}}}f_{\tilde{\alpha}_{\ell}}(\eta^{(\ell)})\,d\eta^{(\ell)}\label{breakint1a} \\
&&+ f_{\beta_{\ell}}(j^{(\ell+1)})\int_{\eta_{j^{(\ell+1)}-1}}^{\eta^{(\ell+1)}}f_{\tilde{\alpha}_{\ell}}(\eta^{(\ell)})\,d\eta^{(\ell)}  \label{breakint1b} \\
&&= \int_{0}^{1}f_{\alpha_{\ell}}(\tilde{\eta}^{(\ell)})\,d\tilde{\eta}^{(\ell)} \sum_{j^{(\ell)}=1}^{j^{(\ell+1)}-1}f_{\beta_{\ell}}(j^{(\ell)})s_{j^{(\ell)}}\label{breakint2a}\\
&&+\int_{0}^{\tilde{\eta}^{(\ell+1)}}f_{\alpha_{\ell}}(\tilde{\eta}^{(\ell)})\,d\tilde{\eta}^{(\ell)} f_{\beta_{\ell}}(j^{(\ell+1)})s_{j^{(\ell+1)}} \label{breakint2b}
\end{eqnarray}
\ees
To obtain Eqs.~\eqref{breakint2a} and \eqref{breakint2b} we rescaled $f_{\tilde{\alpha}_{\ell}}(\eta^{(\ell)})$ in Eq.~\eqref{breakint1a} and \eqref{breakint1b} individually with  
\begin{equation}
\tilde{\eta}^{(\ell)}=\frac{\eta-\eta_{j^{(\ell)}-1}}{s_{j^{(\ell)}}}
\end{equation} 
for  each outer interval $s_{j^{(\ell)}}$, thus obtaining 
$f_{\alpha_{\ell}}(\tilde{\eta}^{(\ell)})$.  In this manner $f_{\alpha_{\ell}}(\tilde{\eta}^{(\ell)})$ is the same function for all the outer intervals,  so that $\int_{0}^{1}f_{\alpha_{\ell}}(\tilde{\eta}^{(\ell)})\,d\tilde{\eta}^{(\ell)}$ can be taken out from the summation, as shown in Eqs.~\eqref{breakint2a} and \eqref{breakint2b}.   

Recall the time-ordering condition, $\eta^{(n)}\geq\eta^{(n-1)}\geq \dots \eta^{(2)}\geq\eta^{(1)}$. It has a consequence that in Eq.~\eqref{breakint2b}, sub-integrals over any two adjacent variables $\eta^{(\ell)}$ and $\eta^{(\ell+1)}$ are nested, as they are in the same outer interval, number $j^{(\ell+1)}$. In this case $\eta^{(\ell)}\leq\eta^{(\ell+1)}$.

In contrast, if the sub-integrals are in different outer intervals (automatically time-ordered), then the sub-integral over $\eta^{(\ell)}$ is not nested inside the subintegral over $\eta^{(\ell+1)}$, but integrated over its entire outer interval independently, as in Eq.  \eqref{breakint2a}.

Let $r_{\ell}=\ast$ denote the time-ordering of outer intervals as in Eq.~\eqref{breakint2a}, and let $r_{\ell}=\emptyset$ denote the integral time-ordering inside a given outer interval as in Eq.~\eqref{breakint2b}. Accordingly, $r_{\ell}$ describes the relation between the adjacent variables  $\eta^{(\ell+1)}$ and $\eta^{(\ell)}$.

As we have just shown, each integral of $a_{\lambda _{n}\dots\lambda _{1}}$ can always be split into  two parts, Eq.~\eqref{breakint2a} and \eqref{breakint2b}, with one exception: if $j^{(\ell+1)}=1$, the subsequent sub-integral of variables $\eta^{(\ell)}$ will only contain the term Eq.~\eqref{breakint2b}. Moreover, both Eq.~\eqref{breakint2a} and \eqref{breakint2b} contain an effective inner part (the part that depends on $f_{\alpha_\ell}$) and an effective outer part (the part that depends on $f_{\beta_\ell}$).  Therefore, by substituting Eqs.~\eqref{breakint2a} and \eqref{breakint2b} into each integral of $a_{\lambda _{n}\dots\lambda _{1}}$, in sequence from $\eta^{(n)}$ to $\eta^{(1)}$, $a_{\lambda _{n}\dots\lambda _{1}}$ can  be written as an inner part $\Phi^{\rm in}$ [Eq.~\eqref{eq:inner}] multiplying an outer part $\Phi^{\rm out}$ [Eq.~\eqref{eq:outer}] over all the possible integration  and summation configurations. Each such configuration can be denoted by an ordered set of symbols $\{r_{n-1}r_{n-2}\dots r_{1}\}$. Thereby, we obtain Eq.~\eqref{form a} as the representation of  $a_{\lambda _{n}\dots\lambda _{1}}$ after this decomposition.


\section{Time symmetry of the UDD pulse intervals}
\label{app:timesym}

Due to the identity $\sin\theta=\sin[\pi-\theta]$, $\sin [\frac{(2j-1)\pi }{2(N+1)}]$ in $s_{j}$ Eq.~\eqref{sj} satisfies 
\begin{eqnarray*}
\sin \frac{(2j-1)\pi }{2(N+1)}&=& \sin[\pi- \frac{(2j-1)\pi }{2(N+1)}%
]  \\
&=& \sin[\frac{(2 N+2-2j+1)\pi }{2(N+1)}] \\
&=& \sin[\frac{(2(N+2-j)-1)\pi }{2(N+1)}]
\end{eqnarray*}
Therefore, we have proved that $s_{j}=s_{N+2-j}$ [Eq.~\eqref{sym}],
which shows that the UDD pulse intervals are time symmetric. There is, however, a difference between even and odd $N$: when $N$ is odd every interval to the left of center is paired with an interval to the right of center. When $N$ is even the central interval is unpaired. E.g., for $N=1$ we have two, paired intervals: $s_1 = s_2$. When $N=2$ we have two paired intervals, $s_1 = s_3$, and an unpaired interval $s_2$.


\section{The parity of the inner order $N_{1}$ determines the parity of $f_{x}$}
\label{app:sys fx}

Since the inner pulse sequences under the piecewise linear variable transformation Eq.~\eqref{linear} still have the UDD$_{N_{1}}$ structure, the rescaled inner pulse intervals remain time symmetric:  
\beq
\theta_{j,k}-\theta_{j,k-1}=\theta_{j,N_{1}+2-k}-\theta_{j,N_{1}+1-k}.
\eeq

When the inner decoupling order $N_{1}$ is even, the parities of $N_{1}+2-k$ and $k$ are the same, so that
\bea
f_{x}(\theta )&=&\begin{cases}
               (-1)^{k-1} & \theta\in [\theta_{j,k-1},\theta _{j,k})\\
               (-1)^{N_{1}+2-k-1} & \theta\in [\theta _{j,N_{1}+1-k},\theta _{j,N_{1}+2-k})
               \end{cases}
               \label{f1}
\eea
Hence $f_{x}(\theta )$ is even inside each outer interval.

When the inner decoupling order $N_{1}$ is odd, 
\bea
f_{x}(\theta )&=&\begin{cases}
               (-1)^{k-1} & \theta\in [\theta_{j,k-1},\theta _{j,k})\\
               (-1)^{N_{1}+2-k} & \theta\in [\theta _{j,N_{1}+1-k},\theta _{j,N_{1}+2-k})
               \end{cases}
               \label{f2}
\eea
where the sign difference between the second lines of Eq.~\eqref{f1} and Eq.~\eqref{f2} arises from the opposite parities of $N_{1}+2-k$ and $k$. Accordingly, $f_{x}(\theta )$ is odd inside each outer interval.

Note that the sequence of rescaled inner intervals $\{\theta_{j,k}\}_{k=1}^{N_1+1}$ is repeated for all values of $j\in\{1,\dots,N_2+1\}$. As a result the three modulation functions $f_x(\theta),f_y(\theta),f_z(\theta)$ are periodic, with respective periods $\frac{\pi}{N_2+1},\frac{2\pi}{N_2+1},\frac{2\pi}{N_2+1}$.
In this sense, the variable transformation $\eta=\sin^{2}(\theta/2)$ introduced in \cite{YangLiu:08}, which emerges naturally from the time structure of UDD sequence Eq.~\eqref{udd time}, is unsuitable for our QDD proof. The reason is that despite the fact that the outer $X$-type pulses intervals are rescaled to be equal, the timing patterns of the inner sequences in different outer intervals are no longer the same.


\section{Fourier expansions of $G(\theta)$}
\label{app:G}

$G(\theta)$ in Eq.~(\ref{G}) takes the following form up to a multiplicative constant: $G(\theta )=s_{j}$, where $\theta \in [\frac{(j-1)\pi }{N+1},\frac{j\pi }{N+1})$. The symmetry property~\eqref{sym} implies that $G(\theta )$ can be written as 
\begin{equation}
G(\theta )=\sum_{\ell =1}^{\infty }g_{\ell }\sin \ell \theta .
\end{equation}
Let us now compute the expansion coefficients:
\begin{eqnarray}
g_{\ell } &\equiv &\frac{1}{\pi /2}\int_{0}^{\pi }G(\theta )\sin \ell \theta d\theta \notag\\
&=&\frac{2}{\pi }\sum_{j=1}^{N+1}s_{j}\int_{\theta _{j-1}}^{\theta _{j}}\sin \ell \theta d\theta \notag\\
&=&-\frac{2}{\pi \ell }\sin \frac{\pi }{2(N+1)}\sum_{j=1}^{N+1}\sin \frac{%
(2j-1)\pi }{2(N+1)} \times \notag \\
&&(\cos \ell \theta _{j}-\cos \ell \theta _{j-1}) \notag\\
&=&-\frac{2}{\pi \ell }\sin \frac{\pi }{2(N+1)} \sum_{j=1}^{N+1}\sin \frac{(2j-1)\pi }{2(N+1)}\times \notag\\
&&(-2)\sin \ell \frac{\theta _{j}+\theta _{j-1}}{2}\sin\ell \frac{\theta _{j}-\theta _{j-1}}{2}
\end{eqnarray}%
where we used the sum-to product formula in the third equality. Due to $\frac{%
\theta _{j}-\theta _{j-1}}{2}=\frac{\pi }{2(N+1)}$ and the product-to sum
formula, we have 
\begin{eqnarray}
g_{\ell }&=&\frac{4}{\pi \ell }\sin \frac{\pi }{2(N+1)}\sin \frac{\ell \pi }{2(N+1)} \times \notag\\
&&\sum_{j=1}^{N+1}\sin \frac{(2j-1)\pi }{2(N+1)}\sin \ell \frac{(2j-1)\pi }{2(N+1)} \notag\\
&=&\frac{4}{\pi \ell }\sin \frac{\pi }{2(N+1)}\sin \frac{\ell \pi }{2(N+1)} \times \\
&&\frac{1}{2}\sum_{j=1}^{N+1}\cos \frac{(\ell -1)(2j-1)\pi }{2(N+1)}-\cos 
\frac{(\ell +1)(2j-1)\pi }{2(N+1)}.\notag
\end{eqnarray}
Considering the sum over $j$ we have
\begin{eqnarray}
&&\sum_{j=1}^{N+1}\cos \frac{(\ell \pm 1)(2j-1)\pi }{2(N+1)}\notag\\
&=&\sum_{j=1}^{N+1}\cos [\frac{(\ell \pm 1)j\pi }{(N+1)}-\frac{(\ell \pm
1)\pi }{2(N+1)}] \notag\\
&=&{\rm Re}[e^{-i\frac{(\ell \pm 1)\pi }{2(N+1)}}\sum_{j=1}^{N+1}e^{i\frac{(\ell
\pm 1)j\pi }{(N+1)}}] \notag \\
&=&{\rm Re}[e^{-i\frac{%
(\ell \pm 1)\pi }{2(N+1)}}e^{i\frac{(\ell \pm 1)\pi }{N+1}}\frac{%
1-e^{i(\ell \pm 1)\pi }}{1-e^{i\frac{(\ell \pm 1)\pi }{N+1}}}] \notag \\
&=&{\rm Re}[\frac{1-\cos (\ell \pm 1)\pi -i\sin (\ell \pm 1)\pi }{e^{-i\frac{(\ell
\pm 1)\pi }{2(N+1)}}-e^{i\frac{(\ell \pm 1)\pi }{2(N+1)}}}] \notag \\
&=&\frac{\sin (\ell \pm 1)\pi }{2\sin \frac{(\ell \pm 1)\pi }{2(N+1)}},
\end{eqnarray}
where in the third equality we used the geometric series formula. The last expression vanishes
if $\ell\neq 2k(N+1)\mp1$. The only values of $\ell$ for which $g_\ell$ does not vanish are $2k(N+1)\mp1$. 
Therefore, finally
\beq
G(\theta )=\sum_{k=0}^{\infty }\sum_{q=\pm1} g_{k,q}\sin
[2k(N+1)\theta +q \theta ] .
\eeq


\end{document}